\documentclass{aa}  
\usepackage{graphicx}
\usepackage{txfonts}
\usepackage{hyperref}
\usepackage{xcolor}
\usepackage[normalem]{ulem}
\begin{document}

   \title{Flux and spectral variability of Mrk\,421 during its moderate activity state using {\it NuSTAR}: Possible accretion disc contribution?}


   \author{Santanu Mondal,
          \inst{1}
          Priyanka Rani,\inst{2,1} 
          C. S. Stalin,\inst{1}
          Sandip K. Chakrabarti\inst{3}
          \and
          Suvendu Rakshit\inst{4}
          }

   \institute{Indian Institute of Astrophysics, II Block, Koramangala, Bangalore 560034, India\\
            \email{santanuicsp@gmail.com}
         \and
             Inter University Centre for Astronomy and Astrophysics, Pune, India
             \and
             Indian Centre for Space Physics, 43 Chalantika, Garia Stn. Road, Kolkata 700084, India
             \and
             Aryabhatta Research Institute of Observational Sciences (ARIES), Nainital 263002, India
             }

 
  \abstract
   {The X-ray emission in BL Lac objects 
is believed to be dominated by synchrotron emission from their 
relativistic jets. However, when the jet emission is not strong, 
one could expect signatures of X-ray emission
from inverse Compton scattering of accretion disc photons by hot and energetic electrons in the corona. Moreover, the observed X-ray variability 
can also originate in the disc, and gets propagated and amplified by the jet.}
   {Here, we present results 
on the BL Lac object Mrk\,421 using the {\it Nuclear Spectroscopic Telescope Array}
data acquired during 2017 when the source was in a moderate X-ray brightness state. For comparison 
with high jet activity state, we also considered one epoch data in April 2013 when the source was in a very high X-ray brightness state. Our aim is to explore the possibility of the signature of accretion disc emission in the overall X-ray emission from Mrk\,421 and also examine changes in accretion parameters considering their contribution to spectral variations.}
   {We divided each epoch of data into different segments in order to find small scale variability. Data for all segments were fitted using simple powerlaw model. We also fitted the full epoch data using the two component advective flow model to extract the accretion flow parameters. Furthermore, we estimated the X-ray flux coming from the different components of the flow using the lowest normalization method and analysed the relations between them.  For consistency, we performed the spectral analysis using available models in the literature.}
   {The simple powerlaw function does not fit the spectra well, and a cutoff needs to be added. The spectral fitting of the data using the two component advective flow model shows that the data can be explained with a model where (a) the size of the dynamic corona at the base of the jet from $\sim 28$ to $10~r_s$, (b) the disc mass accretion rate from 0.021 to 0.051 $\dot M_{\rm Edd}$, (c) the halo mass accretion rate from 0.22 to 0.35 $\dot M_{\rm Edd}$, and (d) the viscosity parameter of the Keplerian accretion disc from 0.18 $-$ 0.25.  In the assumed model, the total flux, disc and jet flux correlate with the radio flux observed during these epochs.}
   {From the spectral analysis, we conclude that the spectra of all the epochs
of Mrk\,421 in 2017 are well described by the accretion disc based two component accretion flow model. The estimated disc and jet flux relations with radio flux show that accretion disc can contribute to the observed X-ray emission, when X-ray data (that covers a small portion of the broad band spectral energy distribution of Mrk\,421) is considered in isolation. 
However, the present disc based models are disfavoured with respect to the relativistic jet models when considering the X-ray data in conjunction with data at other wavelengths.} 

\authorrunning{Mondal et al.}
  \maketitle
\section{Introduction} \label{sec:intro}
Active galactic nuclei (AGN) harbour a supermassive black hole at their center. 
They emit X-rays from their very central region   
and are believed to be powered by accretion of matter
onto supermassive black holes ($\sim 10^6-10^{10} M_\odot$) located at the 
centre. The process of accretion leads to AGN emitting copious 
amount of radiation (luminosity $\sim 10^{42}-10^{48}$~erg s$^{-1}$) as well as 
producing a variety of highly energetic 
phenomena \citep{1969Natur.223..690L,Shakura1973,Rees1984}. A majority of AGN 
emit 
little to very low radio-emission and are called radio-quiet AGN, while a minority emit 
ample radiation in the radio band and are called radio-loud AGN 
\citep{1989AJ.....98.1195K}. However, recently, it has been proposed that AGN 
can be better divided based on their physical differences \citep{Padovani2017} 
into non-jetted  AGN (ones that lack strong relativistic jets) and jetted AGN 
(ones with strong relativistic jets). Among the jetted AGN are blazars, a 
peculiar category characterised by an extreme flux variations across the entire 
electromagnetic spectrum over a range of timescales from minutes to days 
\citep{Wagner1995,Ulrichetal1997}, flat spectra in the radio band, and high 
optical polarization and polarization variability 
\citep{1980ARA&A..18..321A,2005A&A...442...97A,2017ApJ...835..275R}. Such 
characteristics are now attributed to their relativistic jets oriented close 
to the line of sight to the observer and are subjected to Doppler boosting. 

Blazars are divided into two categories, namely, flat spectrum radio quasars 
(FSRQs) and BL Lac objects (BL Lacs). This division is historically based on 
the presence or absence of emission lines in the spectra of those objects. FSRQs 
have emission lines in their optical/infrared spectra, while BL Lacs have either 
weak emission lines (equivalent width $<$ 5 \AA) or featureless spectra 
\citep{UrryPadovani1995}. It is now believed that FSRQs are powerful sources 
with strong relativistic jets and powered by the standard optically thick and 
geometrically thin accretion disc. On the other hand, BL Lacs are weak sources 
with radiatively inefficient and geometrically thin accretion disc. In such discs, 
the reduced ultra-violet (UV) emission  might not be able to photo-ionise the 
broad line region (BLR) leading to weak or absent broad emission lines 
in BL Lacs \citep{Ghisellini2019}.
 
Alternatively, the weakness or absence of emission lines in the spectra of 
BL Lacs could be understood in the context of the continuum getting enhanced 
due to Doppler boosting, and becoming dominant during the increased jet 
activity of BL Lacs. It has been shown by \cite{Foschini2012} that the faint 
lines detected in BL Lacs could be due to the combined effect of the boosted 
jet emission and a weak accretion disc. Removal of the contribution of the 
boosted jet emission to the continuum leads to emission lines in BL Lacs 
having FWHM similar to that of other AGN. This finding leads to hypothesize 
that during the low jet activity or faint state of BL Lacs, the observed 
X-ray emission could have significant contribution from the inverse 
Compton (IC) scattering of soft photons from the accretion disc  off the ``hot corona", 
in addition to 
the synchrotron  emission  from the relativistic jet. According to 
\cite{2011MNRAS.414.2674G} a physical distinction between FSRQs and BL Lacs can 
be made based on the ratio of the luminosity of the BLR to the Eddington 
luminosity ($L_{\rm BLR}/L_{\rm Edd}$) with FSRQs having 
$L_{\rm BLR}/L_{\rm Edd}$ greater than 5 $\times$ $10^{-5}$ compared to 
BL Lacs. 

The broad band spectral energy distribution (SED) of blazars, 
shows two 
prominent peaks, one peaking around radio-optical-X-ray frequencies and the 
other one peaking at X-ray to GeV $\gamma$-ray energies. In the leptonic scenario 
of emission from blazar jets, the low energy peak is attributed to synchrotron 
emission process, while the high energy peak is attributed to IC 
process. Also, based on the location of the synchrotron peak in the SED, 
blazars are further divided in low synchrotron peaked 
(LSP; $\nu_{\rm peak} < 10^{14}$ Hz), intermediate synchroton peaked 
(ISP; $\nu_{\rm peak} = 10^{14-15}$ Hz) and high synchrotorn peaked 
(HSP; $\nu_{\rm peak} > 10^{15}$ Hz) blazars. The observed X-ray emission 
in blazars is mostly by synchrotron or synchrotron self 
Compton (SSC) processes, and thus jet dominated. Therefore, the X-ray flux variations 
in them are believed to be due to their relativistic jets. However, in the 
non-jetted category of AGN such as Seyfert galaxies the observed X-ray 
emission is predominantly due to inverse Compton scattering of optical/UV 
photons from the accretion disc by energetic electrons in the corona 
and thus the X-ray flux variations in them is 
attributed to the accretion disc \citep{HaardtMaraschi1993}. 

In the faint or low jet activity state of FSRQs, prominent accretion disc emission is seen in 
the broad band SED \citep{Raiterietal2009} in many 
sources and therefore in their faint state,  their observed X-ray emission  
could also have a contribution from the accretion disc/corona. Similarly, in the 
case of BL Lacs too, continuum emission is dominated by non-thermal emission 
from their relativistic jets. However, occasionally broad emission lines are 
seen in their optical spectra which suggests that the BLR is 
photoionised by the accretion disc emission. Emission lines (though weak) are 
seen in many BL Lac objects 
\citep{Raiterietal2009,Vermeulen1995,Corbett2000,Stocke2011}. Also, broad 
emission lines are seen in many $\gamma$-ray detected BL Lacs \citep{Shaw2013}, 
a large fraction of them belonging to the LSP and ISP category. Recently, 
from X-ray observations carried out on the BL Lac object Mrk 421 
\cite{Ritabanetal2018} found a break in the power spectral density, which 
the authors attributed 
to the influence of the disc onto the jet emission from Mrk 421. In 
the X-ray light curves of many blazars, log normal (LN) flux distribution is 
observed similar to that seen in Seyfert galaxies 
\citep{Ackermannetal2015,Sinhaetal2016} which again argues for the 
contribution from the accretion disc to the observed X-ray 
emission of blazars. According to \cite{WandelUrry1991}, the observed UV and 
X-ray emission from BL Lacs can come from accretion disc. They were able to 
fit the observed UV and soft X-ray spectrum of the BL Lac object PKS 2155$-$304 
using accretion disc spectra. Also \citet{Grandi2004,ZdziarskiGrandi2001} found the signature of accretion disc emission in the X-ray spectra of the blazar 3C\,273 and rthe adio galaxy 3C\,120. 
However, this model of \cite{WandelUrry1991} was 
not supported by \cite{SmithSitko1991}, who from optical polarimetric 
observations on the BL Lac PKS 2155$-$304, claimed to have found no 
observational evidence for the contribution of the accretion disc to the observed 
UV and optical continuum. A long term multi-wavelength lightcurve and SEDs study showed that the data fits well with the LN model, inferred the lognormality character might have originated from the disc and amplified by the jet \citep{Kapanadze2020,MAGIC2021MNRAS.504.1427A}.

Mrk 421 (z = 0.03 \citealt{deVaucouleursetal1991}) is one of the closest BL Lac objects
hosted in an elliptical galaxy at a distance of 140 Mpc 
($H_0$= 71~km~$s^{-1}$~$Mpc^{-1}$, $\Omega_m$ =0.27, $\Omega_{\Lambda}$ = 0.73)
and hosting a 4$\times$ $10^8$ $M_\odot$ black hole (BH) at the 
center \citep{Wagner2008}.  The SED of Mrk 421 has the classical 
two-peak shape \citep{UrryPadovani1995,Ulrichetal1997} and it has been studied 
extensively in a broad energy band ranging from radio to very high energy 
$\gamma$-rays \citep{Fossatietal2008,Abdoetal2011,Shuklaetal2012,Aleksicetal2015}. Most of 
the observed properties of Mrk 421 are understood to arise from a 
relativistic jet at a small angle along our line of sight 
\citep{UrryPadovani1995}. There are frequent multiwavelength (MWL) campaigns on 
this object to understand its multiscale variability \citep[][and references 
therein]{Macombetal1995,Guptaetal2004,Acciari2009,Paliyaetal2015,Pandeyetal2017}. 
It is an HSP BL Lac and is also the first extra-galactic source detected in 
TeV $\gamma$-rays \citep{Punch1992}. As suggested by several studies, the X-ray
 spectra of HBLs are curved and described well with the log-parabola model 
\citep{Massaroetal2004,Tramacereetal2007}, where the photon index is not a 
constant, rather varies with logarithmic energy. In 2006, the source was 
observed with a peak flux of $\sim 85$~mCrab in the 2$-$10 keV band, 
indicating that the first peak of SED occurred at an energy beyond 10 keV 
\citep{Tramacereetal2009,Ushioetal2009}. Again, in 2013 April, Mrk 421 was 
observed to undergo a major X-ray outburst and was extensively studied by 
multiple observational facilities, including the {\it Nuclear Spectroscopic 
Telescope Array (NuSTAR)} and {\it Swift} 
satellites. Mrk421 showed very high optical emission during its historically low X-ray
emission during January-February 2013 and the presence of multiple compact regions contributing to the broadband emission during low-activity states was suggested by \citet{Balokovicetal2016}. On the contrary, low optical emission during high X-ray
emission was also reported during in February 2010 by \citep{Abeysekaraetal2020}. Therefore, the optical and X-ray emission are found to show different behaviour \citep{Carnereroetal2017}. A MWL study has been performed using the radio to TeV energy band data, very recently, where the authors studied (anti)correlation between optical and hard X-ray energy bands \citep{MAGICmrk4212017data}. Also, it has been studied for its X-ray and $\gamma$-ray flux variability characteristics 
\citep{Kapanadzeetal2016,Ranietal2017,Ranietal2019,Rajput2020}. 

A huge amount of MWL data available on BL Lacs are well 
explained assuming the centre is viewed along the jet. In such a view, though  the jet 
emission significantly gets Doppler boosted, one can also see the central 
region of BL Lacs. Therefore, it is quite natural to expect contributions of 
both the accretion disc in the UV/optical emission and corona 
in the observed X-rays. When the jet emission dominates, 
it might not be possible to find the signature of the accretion disc/corona in the continuum emission from BL Lacs. However, when the jet is less active, emission from accretion 
disc/corona could 
form a significant component of the observed continuum emission. Weak 
(equivalent width $<$ 1 \AA) and narrow FWHM = $300\pm30$ km/sec) Ly$\alpha$ 
emission line has been seen in Mrk\,421 by \cite{Stocke2011}. The weakness of 
the Ly$\alpha$ line in Mrk 421 could be due to the combined effects of (a) the 
Doppler boosting of the continuum emission by the relativistic jet that is 
closely aligned with the observer and (b) weak accretion disc. Taking Doppler 
boosting into account, the intrinsic FWHM of Ly$\alpha$ line in 
Mrk 421 was found to be similar to other AGN \citep{Foschini2012}. These 
observations clearly demonstrate that during the faint state of BL Lacs, the 
accretion disc/corona emission too can be an important contributor to the 
observed X-ray emission, however, might not dominate the non-thermal emission from the jet. In a detailed multiwavelength study by \citet{MAGIC2021MNRAS.504.1427A} showed that the flux distributions of Mrk\,421 in radio and soft X-rays are better described with a Gaussian function, while in the optical, hard X-rays, high energy and VHE $\gamma$-rays are preferably described with a LN distribution. Such distribution of flux implies that the emission is being powered by a multiplicative process rather than an additive one.
In addition, \citet{ChakrabartiDiSilve1994}  
mentioned that very high energy $\gamma$-rays from blazars could be 
due to Fermi acceleration taking place at the surface of the CENtrifugal pressure 
supported BOundary Layer (CENBOL) of a thick disc. Here, CENBOL behaves like a 
dynamical corona also a base of the jet \citep[see][]{MondalChakrabarti2021}. 
At this point, it is worth questing whether a static corona or normal corona
can explain the same behaviour? The answer is no, as the normal corona cannot remain 
hot for ages unless heated by some underlying heating mechanism. It has been observed
that spectral index steepens with age \citep[][and references therein]{Saikiaetal2016}, 
 because jets are not reheated and only 
cools down with age. Therefore, it is essential to consider a dynamical advective halo or
dynamic corona to replenish the electrons. 
Considering the above observed and theoretical constraints, 
in this work, we aimed to fit the 
observed X-ray spectra of Mrk\,421 during its moderately X-ray brightness state with disc based models to derive various disc parameters which could provide useful 
inputs into the accretion characteristics of Mrk\,421. However, we note that in the literature, X-ray emission from Mrk\,421 is explained by processes that do not arise from the disc \citep[][and references therein]{MAGICmrk4212017data}. This paper is organized 
as follows: in Section 2 we explain the observations and data reduction 
procedures, in Section 3, we discuss the various analysis carried out on the 
data sets. The conclusions are given in the final Section.

\section{Observation and data reduction} \label{sec:data_analysis}
\begin{figure*} 
    \centering
	\includegraphics[scale=0.4]{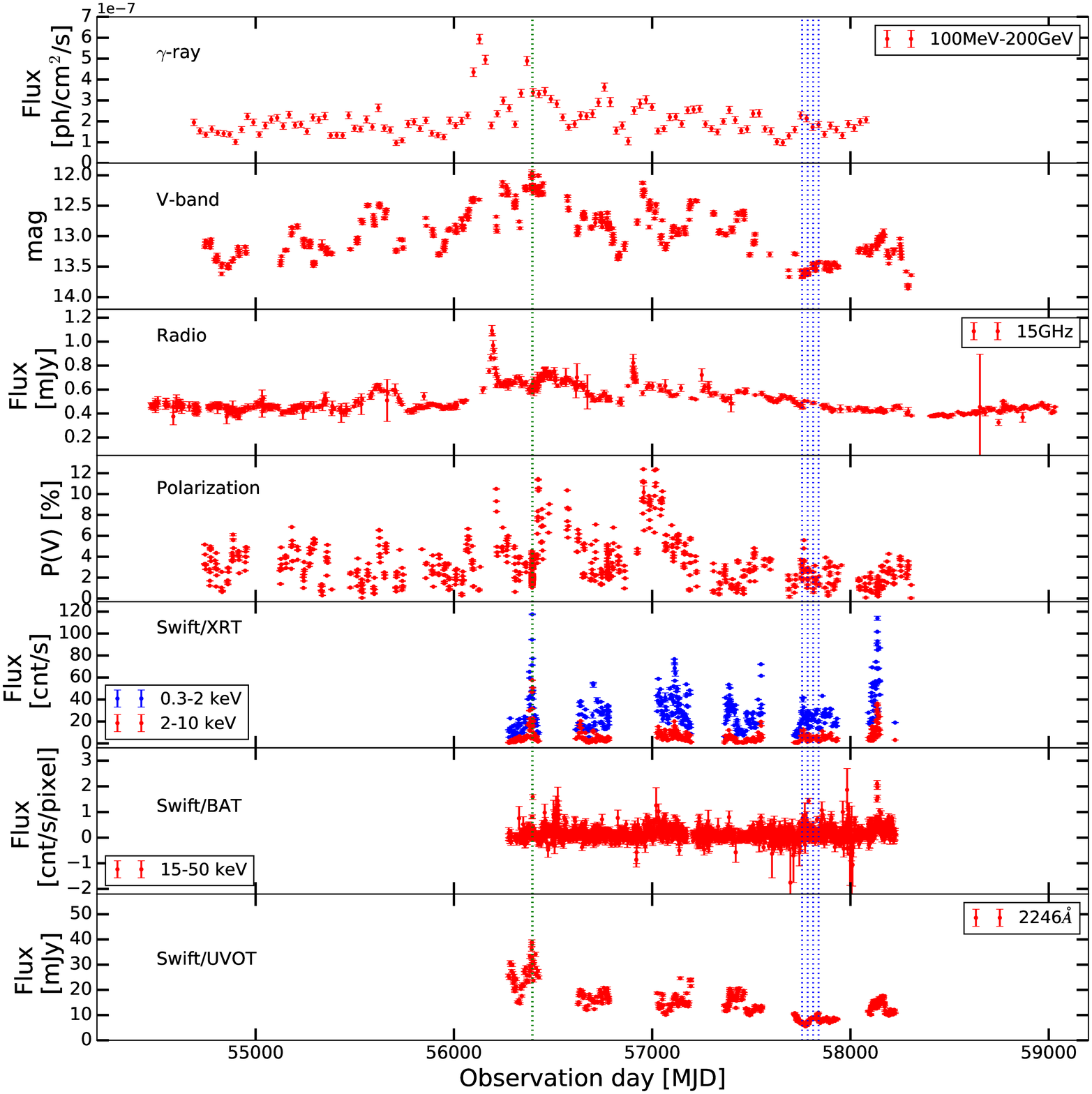}
        \caption{Long term flux and polarization variation of the source.  
From top, we show the flux variation in the $\gamma$-ray, optical and
radio bands. The bottom panel shows the variation in the degree of 
polarization (\%) in the optical V-band.The blue dashed lines show the epochs of {\it NuSTAR} observations studied here. The green dotted line correspond to data taken during April 2013.  
The bottom three panels show the fluxes from the {\it Swift} observation. The X-ray and UV light curves are taken from \citet{Arbet-EngelsEtal2021}. 
} \label{fig:radio}
\end{figure*}

Mrk\,421 was in its moderate brightness state in the radio (a proxy for
jet emission) during 2017 and high brightness state during 2013. 
We show in \autoref{fig:radio}, its MWL
light curves, that include the one month binned $\gamma$-ray light curve for the energy range 100 MeV to 200 GeV \footnote{
https://fermi.gsfc.nasa.gov/ssc/data/access/lat/4yr\_catalog/ap\_lcs.php}
from the {\it Fermi Gamma-ray Space Telescope}, the optical V-band light curve 
from Steward Observatory \citep{2009arXiv0912.3621S} and the 15 GHz radio light curve
from the Owens Valley Radio Observatory (OVRO; \citealt{2011ApJS..194...29R}). The V-band light curve was directly taken from Steward archives and we have not done host galaxy correction to the photometric points.
Also shown in the same figure is the degree of polarization in the optical 
V-band. From the soft X-ray light curve from the {\it Swift/XRT} and the hard X-ray light curve from the {\it Swift/BAT} \citep{Arbet-EngelsEtal2021} it is evident that Mrk\,421 was in a moderate X-ray brightness state during the year 2017 compared to previous years.
The vertical dotted lines in \autoref{fig:radio} show the epochs considered 
in the present paper. Analysis of the X-ray data 
available during the low jet activity state of the source through accretion disc based models fits to the observed X-rays could be a way to derive various accretion disc parameters of the source. We therefore searched the archives of the 
{\it NuSTAR} (\citealt{Harrisonetal2013})  
for the availability of X-ray data and could 
find four epochs of both data available 
during the same period. For comparison, we also selected one epoch during 2013, when the source was in a bright jet activity state that yielded high and variable X-ray and $\gamma$-ray emission \citep{Paliyaetal2015,AcciariEtal2020ApJS}. The details of the observations are given in 
\autoref{table-1}. The flux of the source in the optical, X-ray, and 
$\gamma$-ray bands and the optical polarization during those four epochs are 
given in \autoref{table-2}. From \autoref{table-2}, it is to be noted 
that though the source is 
detected in the $\gamma$-ray band  with good test statistics (TS)
\footnote{A TS of 25 roughly indicates detection at the 
5$\sigma$ level; \cite{1996ApJ...461..396M}} values, it is in a 
relatively quiescent state (see \autoref{fig:radio}). 

Considering the historic brightness state of the source, it is likely that the source is in a relatively faint state in the optical in 2017. \citet{Carnereroetal2017} presented the optical light curves during 2008 - 2015 in R-band. The lowest brightness state was in 2008, when the R-band brightness was ~13.5 mag. The R-band brightness during the four epochs range between 13.09 and 13.32 mag, which suggest that during the four epochs considered in this work, the R-band brightness is close to the historic faint state of Mrk\,421. Low optical emission can always be not a proxy for low X-ray emission as a flare in X-ray can have no corresponding flare in the optical. In such cases, the optical and X-ray emission may not be co-spatial \citep{Carnereroetal2017}. On the contrary, there can be instances of an optical flare with no counterpart in the X-ray band. Also, the multi-band flux variability characteristics of blazars are highly complex, as revealed from recent observations \citet{Rajput2019,Rajput2020,Rajput2021}. However, inspection of \autoref{fig:radio} indicate that the source is in a relatively faint state across different bands of the electromagnetic spectrum during 2017. Thus, during 2017, the contribution of relativistic jet to the total emission in the optical/UV/X-ray may be relatively lower compared to the high jet activity state in 2013.

\begin{figure*} 
    \centering
	\includegraphics[scale=0.4]{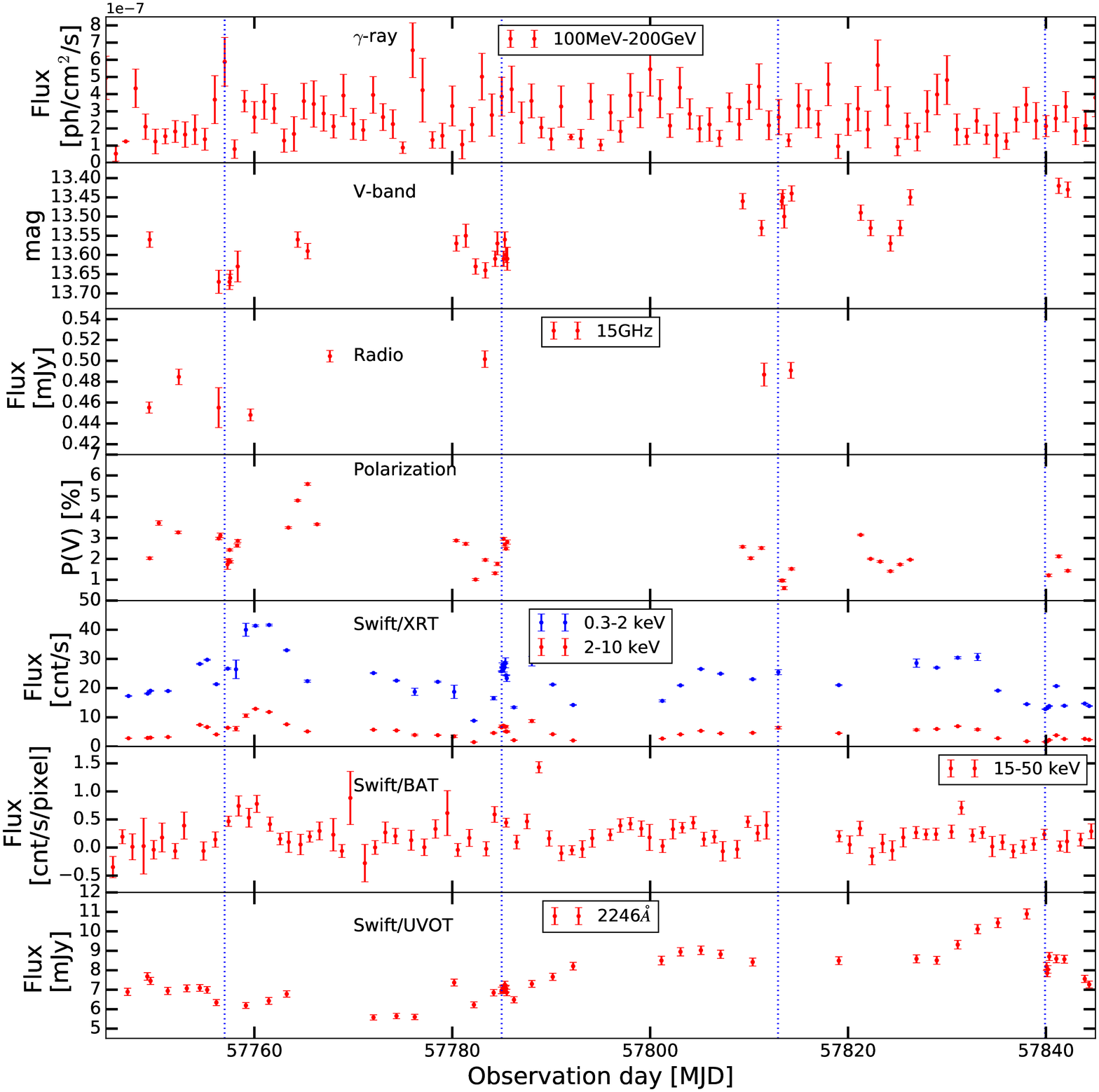}
        \caption{Zoomed version of \autoref{fig:radio}, depicting clearly the brightness states of the four epochs analysed in this work.} \label{fig:ZoomedRadio}
\end{figure*}

The baseline gamma-ray flux (\autoref{fig:radio}) is $\sim 2\times10^{-7}$ ph cm$^{-2}$ s$^{-1}$. During epochs C and D, the gamma-ray activity state is also close to this value. Considering the X-ray light curves from December 2012 - April 2018, the minimum 0.3-2 keV X-ray flux was 5.71 ph s$^{-1}$ and the maximum 0.3-2 keV X-ray flux was 117.44 ph s$^{-1}$. During January - December 2017, the minimum and maximum flux values were 8.84 and 41.64 ph s$^{-1}$ respectively. During the epochs analysed in this work, the flux values ranged between 13.27 and 26.73 ph s$^{-1}$ (\autoref{table-2}). The source is thus not in the faintest X-ray state; however, the X-ray flux values are much weaker than other times and hence the contribution of jet emission during these four epochs is relatively lower compared to other epochs. Also, during the period the $\gamma$-ray, optical and X-ray variations are correlated, and they are also at the moderate brightness states. Between epoch E (historic bright state; fluxes are given in \autoref{sec:jetflux}) and D, the soft flux has decreased by 5.8 times, and the hard (2-10 keV) flux has decreased by 22.6 times. The reduction of hard flux between E and D is much larger, followed by soft X-ray. This points to relatively lower dominance of jet emission in epoch D compared to E and emerging contribution of UV emission from the accretion disc. Between epochs B and D, the hard X-ray flux has decreased by a factor of $\sim 3.5$, while the UV flux has increased by a factor of 1.2. The moderate brightness state in the optical, $\gamma$-ray and radio bands along with low optical polarization in comparison to the polarization at the high X-ray brightness state in 2013 (see \autoref{fig:radio} and \autoref{fig:ZoomedRadio}) indicates that the jet activity in Mrk\,421 during 2017 is likely to be lower.  The $\gamma$-ray lightcurve in \autoref{fig:ZoomedRadio} is one day binned. There are also reports in the literature that point to varied correlation between optical flux and degree of optical polarization \citep{Gauretal2012,PandeyEtal2021}.

\begin{figure*} 

        \includegraphics[height=8.0cm,width=18cm]{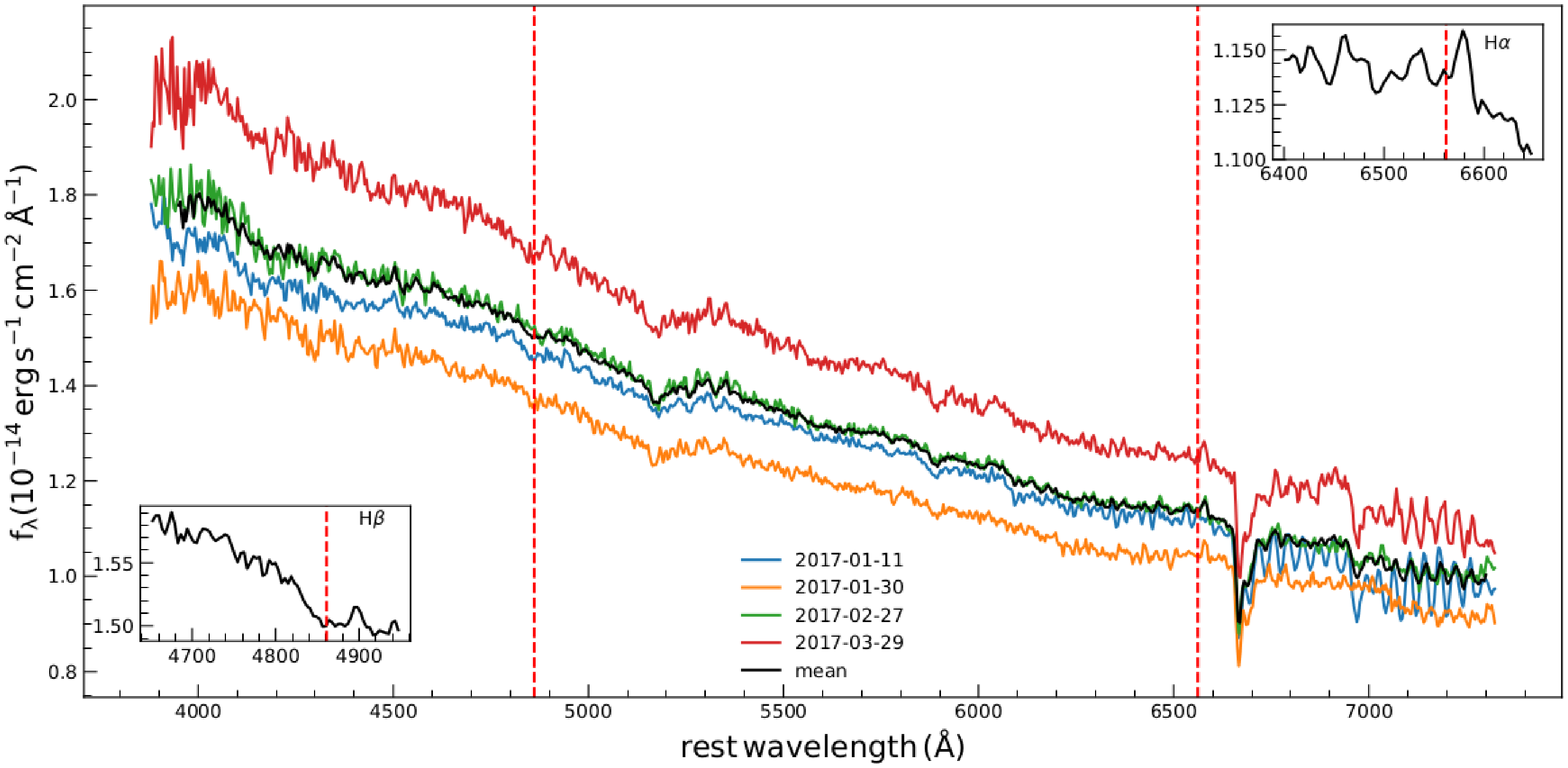}
        \caption{Optical spectra taken from Steward Observatory during the same observation period as considered in {\it NuSTAR} analysis. Four different colors correspond to four epochs considered for X-ray analysis and the black curve shows the combined spectrum. Zoomed regions around H$\alpha$ and H$\beta$ lines are also shown at the top right and bottom left. 
 } \label{fig:OpticalSpectra}
\end{figure*}

Models available in the literature on polarization in blazars point to shock 
propagation through twisted fields inside the
jet being the cause of the changes in polarization \citep[][and references therein]{BlandfordKonigel1979,Lyutikovetal2005,Marscheretal2008,Itohetal2016}. In the present study, of the four epochs considered, in three epochs namely A to C, the source was at a moderate X-ray brightness state, and larger than the typical average X-ray brightness over a quiescent 4-5 months period \citep{Abdoetal2011}. But during epoch D, the X-ray brightness of the source was lower than the typical average quiescent brightness. Also, multi-band SED modelling coupled with other analysis point to the presence of multiple emission zones that give rise to the observed emission even during the quiescent state \citep{Balokovicetal2016}. Such multiple emission regions contributing to the overall observed emission too will give rise to the low level of optical polarisation typically observed in this source, except during instances that are close to flaring episodes. Given the presence of multiple emission regions contributing to the observed broad band emission during the quiescent state of Mrk\,421, we aim to explore the possibility that there is disc emission in the observed X-ray spectrum and under this hypothesis deduce various accretion parameters of the source. We therefore aim to use the Two Component Advective Flow Two Component Advective Flow \citep[{\sc tcaf};][]{ChakrabartiTitarchuk1995} model on the four sets of data analysed in this work. However, we note that multi-band SED fitting of the observations both in quiescent and flaring states invoke SSC processes for the observed X-rays. Recently \citet{Janaetal2017} estimated jet flux from {\sc tcaf} model for systems where the lowest contribution can be treated from the disc only
(and that too not boosted by Doppler effect) while the brightest part will have both the 
disc and the jet. \autoref{fig:OpticalSpectra} shows the optical 
spectra of the source taken from the Steward Observatory
\citep{2009arXiv0912.3621S} archives for the epochs 
considered in this present study. The combined spectrum of the four 
epochs is shown in black. The positions of the H$\alpha$ and H$\beta$ lines are marked 
with vertical dashed lines and zoomed at the upper and lower corners. 
The emission lines are not visible in the optical spectra taken during those four epochs using a moderate size optical telescope. This indicates that even in the moderate X-ray brightness state, the Doppler boosted jet emission dominates to swamp the emission lines. Alternatively, high S/N observations with larger aperture telescopes, might have detected faint emission lines.
The {\it NuSTAR} data were extracted using the standard 
{\sc NUSTARDAS v1.3.1}
\footnote{https://heasarc.gsfc.nasa.gov/docs/nustar/analysis/} software. We 
ran {\sc nupipeline} task to produce cleaned event lists and 
{\sc nuproducts} to generate the spectra. We used a region of 
$30^{\prime\prime}$  for the source and $50^{\prime\prime}$ for the background 
using {\sc ds9}. The data were grouped by {\it grppha} command, with a 
minimum of 30 counts rate per energy bin. The same binning was used for all the 
observations.

As blazar variability timescales can range from 
hours to minutes depending on the underlying physical processes \citep{Aharonianetal2007,Paliyaetal2015}, we split 
each epoch of observation into different segments, with each segment
containing  $\sim1.8$ ksec of data. This division of the data into segments
is to investigate spectral variations in hourly time scale. For that 
purpose, first we made our 
own {\it GTI} files for each time range using the {\it gtibuild} task in 
SAS\footnote{https://www.cosmos.esa.int/web/xmm-newton/sas-threads} 
environment and used those {\it GTI} files during the pipeline extraction 
\citep[same as in][]{MondalChakrabarti2019}. As the data quality in each 
segment is highly noisy above 20~keV, for the analysis of the data pertaining
to each segment we used the data in the energy range of 3$-$20 keV. However, 
considering each epoch of observation as unique, the data is of 
good S/N up to 60 keV, and therefore, for analysis of each epoch of 
observation we used the data in the energy range of 3$-$60 keV. 
For spectral 
analysis of the data, we used {\sc XSPEC}\footnote{https://heasarc.gsfc.nasa.gov/xanadu/xspec/} \citep{Arnaud1996} version 12.11.0. 
Each segment of the data was fitted using the simple power-law ({\sc pl}) model,
while each epoch of observation was fitted using an accretion disc 
based {\sc tcaf} model and the thermal Comptonization model {\sc thcomp} and the {\sc pl} model.
We used the absorption model 
${\it tbabs}$ \citep{Wilmsetal2000} with the Galactic hydrogen column density 
fixed at 1.5$\times$~10$^{20}$~atoms~cm$^{-2}$ \citep{Elvisetal1989,Kalberlaetal2005} throughout the analysis.

\begin{table}
\scriptsize
\centering
\caption{\label{table-1} Log of observations.}
\begin{tabular}{cccccc}
\hline
	OBSID       & Epoch & Date   & MJDstart  &MJDend  &Exposure (s) \\
\hline
60202048002   & A & 03-01-2017 &57756.99385 &57757.52163  & 23691 \\
60202048004   & B & 31-01-2017 &57784.99038 &57785.55288  & 21564  \\
60202048006   & C & 28-02-2017 &57812.92441 &57813.49733  & 23906  \\
60202048008   & D & 27-03-2017 &57839.91052 &57840.63969  & 31228  \\
60002023031   & E & 14-04-2013 &56396.90355 &56397,29939  & 15605 \\
\hline
\end{tabular}
\end{table}

\begin{table*}
\scriptsize
\centering
\caption{\label{table-2} Flux and polarization values of Mrk\,421 during
the epochs studied in this work. The $\gamma$-ray fluxes are in units
of 10$^{-7}$ ph cm$^{-2}$ sec$^{-1}$. The optical flux and polarization
measurements during $^a$ and $^b$ belong to 27 February 2017 and 28 March 2017. The Soft and Hard represent the {\it Swift/XRT} 0.3-2 keV and 2-10 keV band fluxes.}
\begin{tabular}{cccccccccccc}
\hline
	Epoch &Date& $\gamma$-ray  & TS  &Radio & \multicolumn{2}{c}{V-band} &R-band&Soft&Hard&UV&Hard/Soft\\
            &      &  flux       &  &mJy   & (mag) & pol. (\%) &(mag)&ph$s^{-1}$&ph$s^{-1}$&ph$s^{-1}$&\\
\hline
A & 03 Jan. 2017 &3.4 & 97 &$0.46\pm0.02$ & $13.67\pm0.03$&$2.98\pm0.06$&$13.32\pm0.03$&$26.68\pm0.26$&$6.42\pm0.07$&$--$&0.24\\
B & 31 Jan. 2017 &2.1 & 29 &$0.50\pm0.01$ & $13.61\pm0.02$&$1.31\pm0.06$&$13.26\pm0.02$&$26.73\pm1.78$&$6.46\pm0.77$&$7.05\pm0.11$&0.24\\
C & 28 Feb. 2017 &6.9 & 19 &$0.49\pm0.01$ & $13.55\pm0.02^a$&$2.52\pm0.06^a$&$13.19\pm0.03$&$25.48\pm0.85$&$6.46\pm0.42$&$--$&0.25\\
D & 27 Mar. 2017 &2.7 & 11 &$0.48\pm0.01$ & $13.42\pm0.02^b$&$1.43\pm0.05^b$&$13.09\pm0.02$&$13.27\pm0.44$&$1.88\pm0.26$&$8.71\pm0.21$&0.14\\
\hline
\end{tabular}
\end{table*}

\section{Spectral analysis}
\subsection{Model fits to the spectra}
First, we carried out simple {\sc pl} model fits to each segment 
of the data in all the epochs.  
The {\sc tbabs*pl} model fits the data satisfactorily, 
with reduced $\chi_r^2$ between 0.5$-$1.5. The model parameters and 
the reduced $\chi_r^2$ for all the segments are given in \autoref{table:segsPL}. 
Secondly, for all the data in each epoch of observations, we extracted the 
spectra in the 3$-$60~keV band and fitted them with both accretion disc 
and thermal Comptonization models.

For the data belonging to each epoch of {\it NuSTAR}, fitting the spectra
with the phenomenological {\sc pl} model ({\sc tbabs*pl}), was
found not to be satisfactory with $\chi_r^2 > 1.2$ (see
\autoref{table-4}), except for epoch D, where it was unity. 
We further fit the {\it NuSTAR} data using {\sc cutoffpl} model, which returns a good fit with $\chi_r^2 \sim 1$.
We hypothesize this as a signature of a multiprocess system where disc and jet both can contribute.
It is worth mentioning that considering the MWL observation up to TeV band requires synchrotron process based models to fit and infer the data and variability, which are believed to be originated from the jet \citep[see][for a very recent study]{MAGICmrk4212017data}. Also, in the literature, the X-ray spectrum in blazars are known to derive from the conventional power law form with a tendency of show curvature, which is generally attributed to synchrotron cooling break \citep[see][and references therein]{Gaur2020}.

In the literature about blazars, curved X-ray spectra are common and normally used whenever the X-ray instrumental resolution allows to detect such curvature\citep[see e.g.,][]{Massaroetal2004}. Here we provide a different view based on the discussions in the literature (see earlier sections) that variability originated in the disc can be amplified by the jet. Hence, following our hypothesis, where the accretion disc-based model can be applied to Mrk\,421 data during its low or moderate jet activity state.
We therefore used the thermal Comptonization model {\sc thcomp} 
\citep{2020MNRAS.492.5234Z} along with partial ionization {\sc zxipcf} 
\citep{Reeves2008} and {\sc diskbb} as 
{\sc tbabs*zxipcf*thcomp*diskbb} to fit the {\it NuSTAR} data, which returned good fits. 
For proper representation and estimation of parameters of the {\sc thcomp} model,
we used ``energies" command to extend the energy range from 0.01 to 1000 keV.
During the fitting we kept the ionization 
rate (log $\xi$) to 0.0 for the {\sc zxipcf} model and the scattering 
fraction ($f_{\rm sc}$) was kept fixed at 1 for all epochs for the {\sc thcomp} 
model. Addition of {\sc diskbb} provided the temperature of the disc 
($T_{\rm in}$) which was found to be reasonable (see \autoref{table-5}). For 
epoch D, the partial covering factor ($C_f$) was the lowest $0.07$ and the temperature
of the corona was the lowest among the four epochs. This implies 
increased generation of disc photons, which on getting inverse
Compton scattered by the hot 
electrons in the corona cooled the corona down. This
is also consistent with the lowest-flux state of the source at epoch D. 
We found temperature of the corona with values of 
144$\pm$46, 55$\pm$8, 51$\pm$9, and 27$\pm$5 keV for epochs A, B, C,
and D respectively, thus decreasing. 
We thus found the temperature of the corona to decrease from epoch A to D.
The results of the best fitted model parameters are given in \autoref{table-5}. From the {\sc diskbb} model fitted norm we estimated $R_{\rm in}$/$r_{\rm s}$, which comes out to be 57, 25, 17, and 2 for epochs A to D. The estimated values are significant and as per our hypothesis, this could be a signature of the presence of possibly a truncated disc component. 

\begin{figure*}
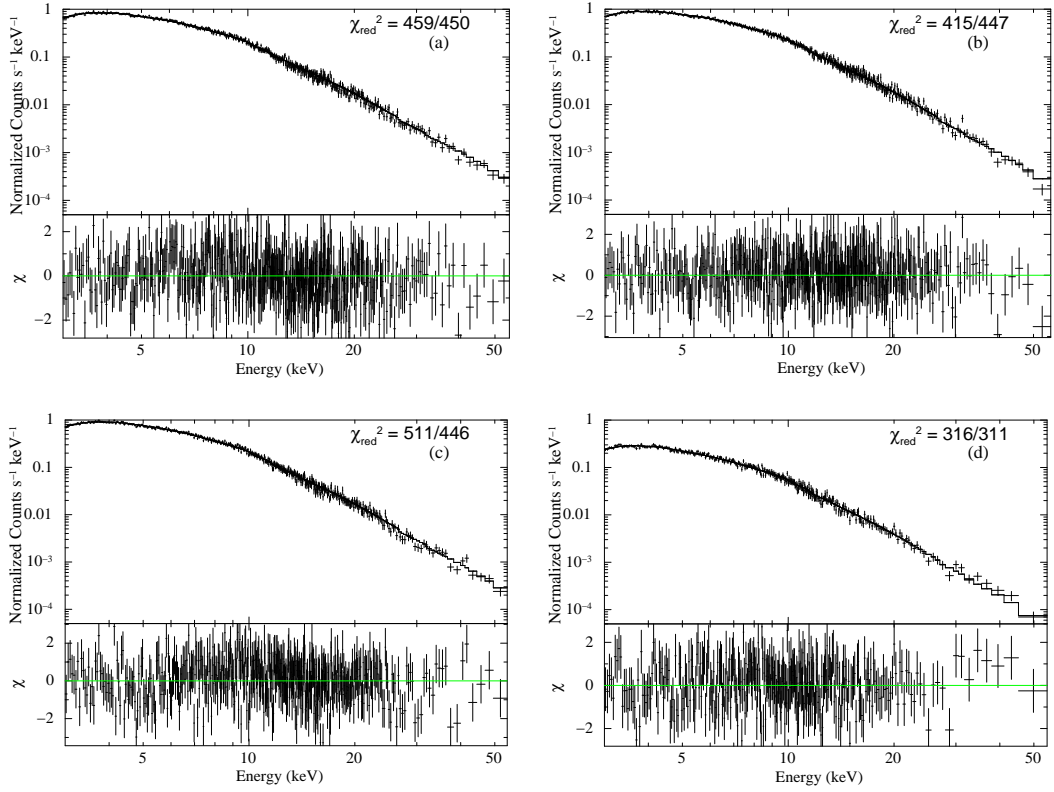
 
    \centering{
    \includegraphics[height=7truecm,angle=270]{Figures/x02-zxipcf-tcaf-R1.eps}
    \includegraphics[height=7truecm,angle=270]{Figures/x04-zxipcf-tcaf-R1.eps}}
    \centering{
  \includegraphics[height=7truecm,angle=270]{Figures/x06-zxipcf-tcaf-R1.eps}
  \includegraphics[height=7truecm,angle=270]{Figures/x08-zxipcf-tcaf-R1.eps}}
\caption{The 3$-$60 keV spectra fitted with {\sc tbabs*zxipcf*tcaf} model along with the residuals. The panels (a)-(d) correspond to the data acquired at epochs A, B, C and D respectively.}
    \label{fig:spectcafpl}
\end{figure*}

Further, to study the accretion flow parameters and 
geometrical variation of the flow and its 
contribution to the spectrum, we fitted the four observations in the 3$-$60 keV band using 
the physical {\sc tcaf} model. For 
spectral fitting, we ran the source code directly in XSPEC as a local 
model\footnote{https://heasarc.gsfc.nasa.gov/xanadu/xspec/manual/node101.html}.
The {\sc tcaf} model\footnote{Currently the model is not available in XSPEC as a local/table model.} \citep{ChakrabartiTitarchuk1995} can successfully fit the X-ray data on low mass X-ray 
binaries (\citealt[][and references therein]{Debnathetal2014,Mondaletal2014,Iyeretal2015,Debnathetal2015}) and AGN \citep{MandalChakrabarti2008,Nandietal2019,MondalStalin2021}, and has been
used to infer the outburst behaviour as well as the variability properties of 
compact objects. This model requires five 
parameters, namely, (i) mass of the black hole ($M_{\rm BH}$), (ii) disc accretion 
rate ($\dot m_d$), (iii) halo accretion rate ($\dot m_h$), (iv) shock 
location, which is the boundary of the CENBOL (\citet{Chakrabarti1989}, $X_s$ in unit of $r_s$, where $r_s$ is the 
Schwarzschild radius, $2GM_{\rm BH}/c^2$), 
and (v) shock compression ratio ($R$). For {\sc tcaf} model fits, we froze the 
mass of the BH at $4\times 10^8$~M$_\odot$ \citep{Wagner2008}, which gives 
$r_s=1.2\times 10^{14} \text{cm}$. The quality of fits for all epochs 
appear to be good with the reduced $\chi_r^2$ ranging from 0.9 to 1.1.
It can be seen that the accretion rate of the disc component varies in the 
range from 0.02$-$0.05$\dot M_{\rm Edd}$ and the halo accretion rate varies 
from 0.22$-$0.35$\dot M_{\rm Edd}$. Increase of the disc accretion rate leads to 
increase in the number of soft photons. This cools the corona by inverse 
Compton scattering, thus causing the inner edge of the disc to move 
inward from $\sim 20$ to $10 r_s$. 
The results of the {\sc tbabs*zxipcf*tcaf} model fits are given in  
\autoref{fig:spectcafpl}. 
From the parameters obtained
from {\sc tbabs*zxipcf*tcaf} fits to the data we derived post-facto values of 
$kT_e$ of 189, 55, 38, and 20 keV for epochs A, B, C, and D respectively. We are not using any simple equation to calculate $kT_e$ in the source code, which depends on all input parameters. Therefore, estimating error is beyond the scope of this paper. 

Thus, from the fitting of four epochs of data with both {\sc thcomp} 
and {\sc tcaf} models we infer the following (i) {\sc tcaf} model alone is 
a good representation of the observations in the full energy band, proving different accretion flow parameters,
(ii) the thermal Comptonization
{\sc thcomp} model along with {\sc diskbb} well fits the data
providing values of the temperature the corona, photon index, and 
the temperature of the disc, (iii) the nature of variation
of $kT_e$ is in agreement with the variation in mass accretion rate, 
(iv) the highest accretion rate obtained for the epoch D from {\sc tcaf} is in 
agreement with the lowest flux, and shock compression ratio observed,
(v) the adequacy of non-magnetic disc models to fit the data, can be the signature of low magnetic field.

\subsection{Photon index and spectral flux}
\begin{figure*}[!h]
    \centering{
    \includegraphics[height=6truecm,angle=0]{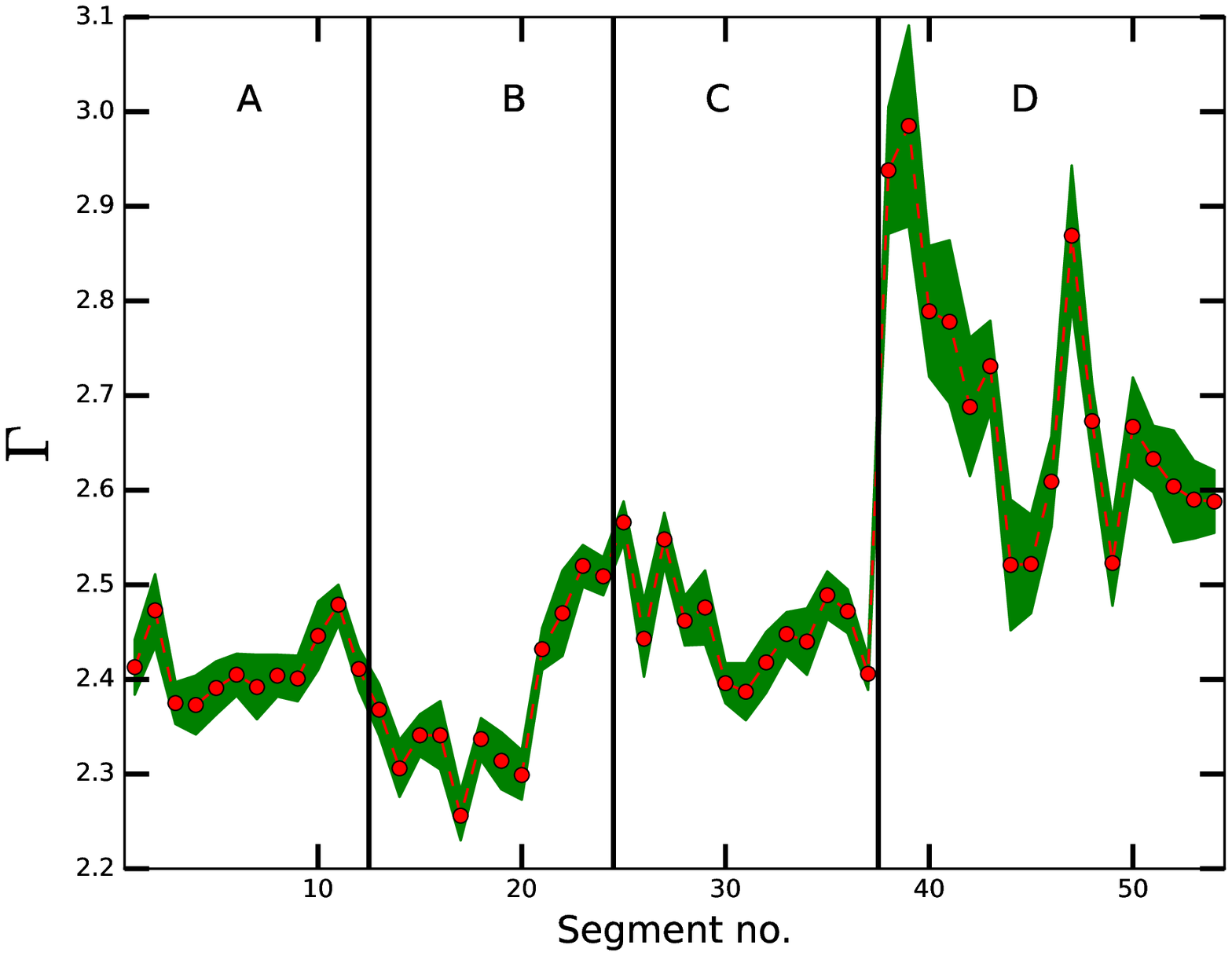}
    \hspace{-0.5cm}
    \includegraphics[height=6truecm,angle=0]{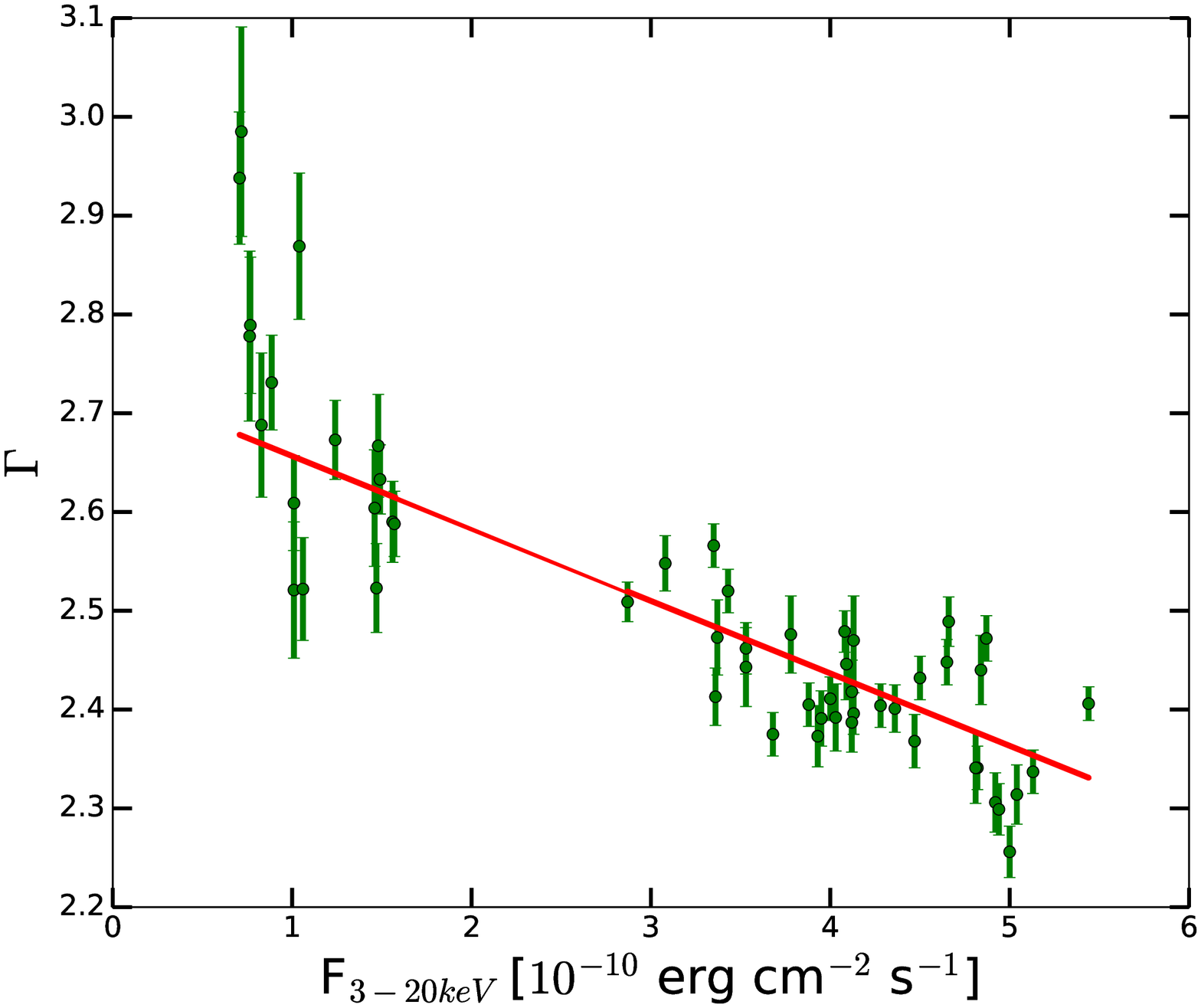}}
    \caption{Variation of $\Gamma_{\rm PL}$ with segment number (left panel). 
The green shaded band is the one $\sigma$ error region and the black lines 
show the transition from one observation to the other. The right panel shows 
the $\Gamma_{\rm PL}$ variation with total flux. The  red solid line is the weighted 
linear least square fit to the data. Both panels show a significant change 
in $\Gamma_{\rm PL}$ from 2.2 to 3.0 in about 83~days.}
    \label{fig:PLindexFlux}
\end{figure*}

On hourly time scales, the photon index ($\Gamma_{\rm PL}$) is found to vary 
between $2.2$ and $3.0$ ($\Delta \Gamma=0.8 $)
within a span of 83~days from epoch A to D. 
We note that $\Gamma_{\rm PL}$ was always high, which implies 
that the source was in a high/soft spectral state. The very low flux spectrum 
with high $\Gamma_{\rm PL}$ belongs to epoch D. 
\autoref{fig:PLindexFlux}, left panel shows the variation of 
$\Gamma_{\rm PL}$ with segments of every epoch of observation. 
Significant change in $\Gamma_{\rm PL}$ was observed during epoch D. 
On each epoch of observation, the source exhibited small scale fluctuations in $\Gamma_{\rm PL}$. 
The green shaded band shows the error bar and black vertical lines show 
transition in each epoch. To check the behaviour of $\Gamma_{\rm PL}$ of all the 
segments with the total flux in the 3$-$20 keV band, we plotted the variation 
of $\Gamma_{\rm PL}$ with the total flux $F_{3-20~keV}$ in \autoref{fig:PLindexFlux}. 
We fitted the observed data points in the $\Gamma$ vs.
flux diagram using a linear function of the form 
$\Gamma = a \times F_{3-20keV} + b$, taking into account the errors in 
both $\Gamma$ and flux following \cite{1992nrca.book.....P}. This fitting gave 
a significant negative correlation between $\Gamma$ and flux with a 
correlation co-efficient of $-$0.860, i.e. the value of photon index decreases 
with increasing flux indicating a `harder when brighter' trend in the 3$-$20 keV band. The same behaviour has also been observed using the same data by \citet{MAGICmrk4212017data}. 
Such behaviour is more often seen in the HSP category of 
blazars \citep{Giommietal1990,Pianetal1998}. Mrk~421  is an HSP blazar
and the harder when brighter trend could be most likely due to change in the 
power-law component of the relativistic jet \citep{Ranietal2017}.
However, such harder when brighter trend in the X-ray band can also be due 
to processes related to the accretion disc /corona as discussed towards
the end of the next section. 
\begin{table*}
\scriptsize
\centering
\caption{\label{table:segsPL} Results of {\sc tbabs*pl} model fits to each 
segment from epoch A to D. The fluxes are in units of $10^{-10} \text{erg} \text{cm}^{-2} \text s^{-1}$. Here, $\Gamma_{\rm PL}$ is the powerlaw photon index.}
\begin{tabular}{cccccccccc}
\hline
Epoch & Seg. No.&MJDstart &MJDend &$\Gamma_{\rm PL}$ &$F_{3-10\text{keV}}$ & $F_{10-20\text{keV}}$&$F_{3-20\text{keV}}$&HR&$\chi_r^2/dof$\\
\hline
  &  1 &57756.99385 &57757.03783 & $2.413\pm  0.029$ & $2.423\pm  0.003$ & $0.937\pm  0.001$ & $3.360\pm  0.006$ &$0.387\pm  0.001$&1.0/158  \\ 
  &  2 &57757.03783 &57757.08181 & $2.473\pm  0.038$ & $2.472\pm  0.005$ & $0.901\pm  0.001$ & $3.374\pm  0.012$ &$0.364\pm  0.001$&0.8/99   \\
  &  3 &57757.08181 &57757.12580 & $2.375\pm  0.022$ & $2.623\pm  0.001$ & $1.053\pm  0.014$ & $3.676\pm  0.005$ &$0.401\pm  0.005$&1.2/208  \\
  &  4 &57757.12580 &57757.16978 & $2.373\pm  0.031$ & $2.805\pm  0.003$ & $1.128\pm  0.008$ & $3.933\pm  0.005$ &$0.402\pm  0.003$&0.9/154  \\
  &  5 &57757.16978 &57757.21376 & $2.391\pm  0.028$ & $2.829\pm  0.001$ & $1.118\pm  0.001$ & $3.947\pm  0.005$ &$0.395\pm  0.001$&0.9/161  \\
A &  6 &57757.21376 &57757.25774 & $2.405\pm  0.022$ & $2.793\pm  0.008$ & $1.089\pm  0.004$ & $3.883\pm  0.005$ &$0.390\pm  0.002$&1.0/207  \\
  &  7 &57757.25774 &57757.30172 & $2.392\pm  0.034$ & $2.889\pm  0.009$ & $1.140\pm  0.005$ & $4.029\pm  0.002$ &$0.395\pm  0.002$&1.0/123  \\
  &  8 &57757.30172 &57757.34570 & $2.404\pm  0.022$ & $3.075\pm  0.008$ & $1.200\pm  0.007$ & $4.275\pm  0.009$ &$0.390\pm  0.002$&1.0/202  \\
  &  9 &57757.34570 &57757.38968 & $2.401\pm  0.024$ & $3.131\pm  0.002$ & $1.224\pm  0.006$ & $4.355\pm  0.011$ &$0.391\pm  0.002$&0.9/186  \\
  & 10 &57757.38968 &57757.43367 & $2.446\pm  0.036$ & $2.977\pm  0.010$ & $1.114\pm  0.013$ & $4.091\pm  0.002$ &$0.374\pm  0.005$&0.8/123  \\
  & 11 &57757.43367 &57757.47765 & $2.479\pm  0.021$ & $2.992\pm  0.007$ & $1.084\pm  0.006$ & $4.077\pm  0.011$ &$0.362\pm  0.002$&1.2/213  \\
  & 12 &57757.47765 &57757.52163 & $2.411\pm  0.022$ & $2.883\pm  0.009$ & $1.117\pm  0.004$ & $4.000\pm  0.005$ &$0.387\pm  0.002$&1.0/209  \\
\hline
  & 13 &57784.99038 &57785.03725 & $2.368\pm  0.027$ & $3.186\pm  0.002$ & $1.287\pm  0.005$ & $4.473\pm  0.004$ &$0.404\pm  0.002$&1.2/181  \\
  & 14 &57785.03725 &57785.08413 & $2.306\pm  0.030$ & $3.439\pm  0.015$ & $1.477\pm  0.004$ & $4.916\pm  0.004$ &$0.429\pm  0.002$&0.9/146  \\
  & 15 &57785.08413 &57785.13100 & $2.341\pm  0.022$ & $3.408\pm  0.005$ & $1.415\pm  0.005$ & $4.824\pm  0.002$ &$0.415\pm  0.002$&1.0/199  \\
  & 16 &57785.13100 &57785.17788 & $2.341\pm  0.036$ & $3.403\pm  0.008$ & $1.411\pm  0.009$ & $4.814\pm  0.003$ &$0.415\pm  0.003$&0.9/106  \\
  & 17 &57785.17788 &57785.22475 & $2.256\pm  0.026$ & $3.440\pm  0.007$ & $1.549\pm  0.004$ & $5.000\pm  0.004$ &$0.450\pm  0.001$&1.1/173  \\
  & 18 &57785.22475 &57785.27163 & $2.337\pm  0.022$ & $3.619\pm  0.004$ & $1.507\pm  0.014$ & $5.126\pm  0.010$ &$0.416\pm  0.004$&1.1/208  \\
B & 19 &57785.27163 &57785.31850 & $2.314\pm  0.030$ & $3.537\pm  0.009$ & $1.507\pm  0.006$ & $5.045\pm  0.005$ &$0.426\pm  0.002$&0.8/154  \\
  & 20 &57785.31850 &57785.36538 & $2.299\pm  0.026$ & $3.449\pm  0.005$ & $1.491\pm  0.005$ & $4.939\pm  0.004$ &$0.432\pm  0.002$&1.0/177  \\
  & 21 &57785.36538 &57785.41225 & $2.432\pm  0.022$ & $3.259\pm  0.010$ & $1.237\pm  0.006$ & $4.496\pm  0.001$ &$0.380\pm  0.002$&1.0/199  \\
  & 22 &57785.41225 &57785.45913 & $2.470\pm  0.045$ & $3.025\pm  0.012$ & $1.105\pm  0.010$ & $4.130\pm  0.023$ &$0.365\pm  0.004$&0.8/75   \\
  & 23 &57785.45913 &57785.50600 & $2.520\pm  0.022$ & $2.542\pm  0.004$ & $0.884\pm  0.003$ & $3.426\pm  0.018$ &$0.348\pm  0.001$&1.1/200  \\
  & 24 &57785.50600 &57785.55288 & $2.509\pm  0.020$ & $2.127\pm  0.002$ & $0.747\pm  0.002$ & $2.874\pm  0.008$ &$0.351\pm  0.001$&1.1/241  \\
\hline
 &  25 &57812.92441 &57812.96848 & $2.566\pm  0.022$ & $2.510\pm  0.005$ & $0.834\pm  0.003$ & $3.345\pm  0.011$ &$0.332\pm  0.001$&1.0/189  \\
 &  26 &57812.96848 &57813.01255 & $2.443\pm  0.040$ & $2.565\pm  0.003$ & $0.963\pm  0.007$ & $3.528\pm  0.014$ &$0.375\pm  0.003$&0.6/86   \\
 &  27 &57813.01255 &57813.05662 & $2.548\pm  0.028$ & $2.301\pm  0.001$ & $0.775\pm  0.005$ & $3.080\pm  0.006$ &$0.337\pm  0.002$&0.7/162  \\
 &  28 &57813.05662 &57813.10069 & $2.462\pm  0.026$ & $2.577\pm  0.003$ & $0.950\pm  0.001$ & $3.526\pm  0.004$ &$0.369\pm  0.001$&0.9/171  \\ 
 &  29 &57813.10069 &57813.14476 & $2.476\pm  0.039$ & $2.773\pm  0.018$ & $1.008\pm  0.002$ & $3.781\pm  0.022$ &$0.364\pm  0.002$&0.9/98   \\
 &  30 &57813.14476 &57813.18883 & $2.396\pm  0.021$ & $2.968\pm  0.008$ & $1.166\pm  0.005$ & $4.134\pm  0.014$ &$0.393\pm  0.002$&1.2/215  \\
 &  31 &57813.18883 &57813.23291 & $2.387\pm  0.030$ & $2.951\pm  0.003$ & $1.170\pm  0.002$ & $4.121\pm  0.009$ &$0.396\pm  0.001$&1.0/148  \\
C&  32 &57813.23291 &57813.27698 & $2.418\pm  0.032$ & $2.978\pm  0.006$ & $1.146\pm  0.002$ & $4.124\pm  0.002$ &$0.385\pm  0.001$&0.9/136  \\
 &  33 &57813.27698 &57813.32105 & $2.448\pm  0.023$ & $3.389\pm  0.002$ & $1.265\pm  0.003$ & $4.654\pm  0.008$ &$0.373\pm  0.001$&1.1/198  \\
 &  34 &57813.32105 &57813.36512 & $2.440\pm  0.035$ & $3.515\pm  0.006$ & $1.323\pm  0.002$ & $4.839\pm  0.031$ &$0.376\pm  0.001$&1.1/118  \\
 &  35 &57813.36512 &57813.40919 & $2.489\pm  0.025$ & $3.432\pm  0.013$ & $1.232\pm  0.001$ & $4.664\pm  0.021$ &$0.359\pm  0.001$&1.1/183  \\
 &  36 &57813.40919 &57813.45326 & $2.472\pm  0.023$ & $3.568\pm  0.009$ & $1.301\pm  0.003$ & $4.869\pm  0.010$ &$0.365\pm  0.001$&0.9/191  \\
 &  37 &57813.45326 &57813.49733 & $2.406\pm  0.017$ & $3.919\pm  0.006$ & $1.525\pm  0.005$ & $5.444\pm  0.017$ &$0.389\pm  0.001$&1.2/244  \\
\hline
  & 38 &57839.91052 &57839.95341 & $2.938\pm  0.067$ & $0.576\pm  0.001$ & $0.131\pm  0.003$ & $0.707\pm  0.002$ &$0.227\pm  0.005$&0.9/39   \\
  & 39 &57839.95341 &57839.99630 & $2.985\pm  0.106$ & $0.589\pm  0.001$ & $0.130\pm  0.005$ & $0.717\pm  0.002$ &$0.221\pm  0.008$&1.5/19   \\
  & 40 &57839.99630 &57840.03920 & $2.789\pm  0.069$ & $0.612\pm  0.004$ & $0.163\pm  0.001$ & $0.767\pm  0.008$ &$0.266\pm  0.002$&1.0/46   \\
  & 41 &57840.03920 &57840.08209 & $2.778\pm  0.086$ & $0.601\pm  0.009$ & $0.161\pm  0.003$ & $0.763\pm  0.001$ &$0.268\pm  0.006$&0.8/28   \\
  & 42 &57840.08209 &57840.12498 & $2.688\pm  0.073$ & $0.640\pm  0.001$ & $0.188\pm  0.002$ & $0.829\pm  0.001$ &$0.294\pm  0.003$&0.8/30   \\
  & 43 &57840.12498 &57840.16787 & $2.731\pm  0.048$ & $0.697\pm  0.004$ & $0.196\pm  0.002$ & $0.886\pm  0.005$ &$0.281\pm  0.003$&0.9/72   \\
  & 44 &57840.16787 &57840.21077 & $2.521\pm  0.069$ & $0.750\pm  0.002$ & $0.261\pm  0.002$ & $1.011\pm  0.008$ &$0.348\pm  0.003$&1.1/37   \\
  & 45 &57840.21077 &57840.25366 & $2.522\pm  0.052$ & $0.785\pm  0.001$ & $0.272\pm  0.001$ & $1.057\pm  0.003$ &$0.346\pm  0.001$&1.1/61   \\
D & 46 &57840.25366 &57840.29655 & $2.609\pm  0.048$ & $0.765\pm  0.006$ & $0.244\pm  0.002$ & $1.009\pm  0.005$ &$0.319\pm  0.004$&1.0/73   \\
  & 47 &57840.29655 &57840.33944 & $2.869\pm  0.074$ & $0.836\pm  0.003$ & $0.205\pm  0.002$ & $1.040\pm  0.009$ &$0.245\pm  0.003$&0.5/129  \\
  & 48 &57840.33944 &57840.38234 & $2.673\pm  0.040$ & $0.951\pm  0.004$ & $0.284\pm  0.003$ & $1.235\pm  0.002$ &$0.299\pm  0.003$&0.9/104 \\
  & 49 &57840.38234 &57840.42523 & $2.523\pm  0.045$ & $1.093\pm  0.005$ & $0.380\pm  0.003$ & $1.472\pm  0.003$ &$0.348\pm  0.003$&1.2/77   \\
  & 50 &57840.42523 &57840.46812 & $2.667\pm  0.052$ & $1.141\pm  0.001$ & $0.343\pm  0.003$ & $1.484\pm  0.002$ &$0.301\pm  0.003$&0.9/59   \\
  & 51 &57840.46812 &57840.51101 & $2.633\pm  0.035$ & $1.138\pm  0.001$ & $0.353\pm  0.004$ & $1.491\pm  0.007$ &$0.310\pm  0.004$&1.2/121  \\
  & 52 &57840.51101 &57840.55391 & $2.604\pm  0.059$ & $1.109\pm  0.001$ & $0.355\pm  0.003$ & $1.464\pm  0.007$ &$0.320\pm  0.003$&0.9/47   \\
  & 53 &57840.55391 &57840.59680 & $2.590\pm  0.041$ & $1.175\pm  0.001$ & $0.381\pm  0.001$ & $1.556\pm  0.394$ &$0.324\pm  0.001$&0.9/88   \\
  & 54 &57840.59680 &57840.63969 & $2.588\pm  0.033$ & $1.186\pm  0.004$ & $0.386\pm  0.002$ & $1.572\pm  0.006$ &$0.325\pm  0.002$&1.0/133  \\
\hline
\end{tabular}
\end{table*}

\begin{table}
\scriptsize
\centering
\caption{Best fitting parameters and fluxes for the models {\sc tbabs*pl} to {\it NuSTAR}. Here, flux is in $10^{-10}$ erg cm$^2$sec$^{-1}$ unit.}
\label{table-4}
\begin{tabular} {cccccc}
\hline
Epoch&MJDstart &MJDend & $\Gamma_{\rm PL}$ & $F_{\rm 3-60 keV}$ &  $\chi_r^2/dof$  \\ 
\hline
A &57756.99385 &57757.52163 &$2.436\pm0.006$  &$5.166\pm0.017$& 1.2/456 \\
B &57784.99038 &57785.55288 &$2.421\pm0.006$  &$5.547\pm0.006$& 1.4/452 \\
C &57812.92441 &57813.49733 &$2.488\pm0.006$  &$5.335\pm0.022$& 1.5/451 \\
D &57839.91052 &57840.63969 &$2.661\pm0.011$  &$1.430\pm0.001$& 1.0/316 \\
\hline
\end{tabular}
\end{table}

\begin{table*}
\scriptsize
\centering
\caption{\label{table-5} Parameters of  
{\sc tbabs*zxipcf*thcomp*diskbb} model fits to the spectra. Here, $C_{\rm f}$ 
is the partial covering fraction, $kT_e$, and $\Gamma$ are the 
temperature of the corona and photon index of the spectrum respectively. 
$T_{\rm in}$ and N$_{\rm dbb}$ are the temperature 
 and normalization of the {\sc diskbb} model component.}
\begin{tabular}{cccccccccc}
\hline
Epoch &MJDstart &MJDend &$C_{\rm f}$  & $kTe$ &  $\Gamma$ & $T_{\rm in}$ &N$_{\rm dbb}$ &$\chi_r^2/dof$ \\
      & &&& [keV] &       &[keV] & $\times 10^{13}$ & \\
\hline
A &57756.99385 &57757.52163 & $0.26\pm0.02$ &$144\pm46$ &$2.49\pm0.06$& $0.07\pm0.02$ &$3.13\pm1.02$ &1.0/451\\
B &57784.99038 &57785.55288 & $0.33\pm0.07$ &$55\pm8$  &$2.53\pm0.09$&$0.08\pm0.02$ &$0.61\pm0.17$&0.9/448\\
C &57812.92441 &57813.49733 & $0.3\pm0.03$ &$51\pm9$  &$2.6\pm0.1$ &$0.08\pm0.02$&$0.26\pm0.05$&1.1/446\\
D &57839.91052 &57840.63969 & $0.07\pm0.01$ &$27\pm5$ &$2.58\pm0.07$ &$0.09\pm0.02$&2.9e$-$3$\pm$3e$-$4&1.0/312\\
\hline
\end{tabular} 
\end{table*}

\begin{table*}
\scriptsize
\centering
\caption{\label{table:tcafPL} Results of {\sc tbabs*zxipcf*tcaf} model fits to the spectra. 
$\dot m_{\rm d}$, and $\dot m_{\rm h}$ represent {\sc tcaf} model fitted sub-Keplerian 
(halo) and Keplerian (disc) rates in Eddington rate unit respectively.  
$X_s$ (in Schwarzchild radius $r_s$), and $R$ are the model fitted shock 
location and shock compression ratio values respectively. $N_t$ is the model normalization. Fluxes are in $10^{-10}$ ergs~cm$^{-2}$~s$^{-1}$.} 
\begin{tabular}{cccccccccccccccc}
\hline
MJDstart  &MJDend &$C_{\rm f}$&$\dot m_{\rm d}$ &$\dot m_{\rm h}$ & $X_{\rm s}$ &  R & $N_t \times 10^4$ &$F_{\rm tot}$&$F_{\rm disk}$&$F_{\rm jet}$&$\chi_r^2$ \\
\hline
57756.99385 &57757.52163 &$0.31\pm0.03$&$0.021\pm0.002$&$ 0.221\pm0.011$ &$19.77\pm3.79$& $1.91\pm0.31$&$1.36\pm0.18$ &$5.09\pm0.02$&$1.61\pm0.01$&$3.48\pm0.02$&1.0/450 \\
57784.99038 &57785.55288 &$0.36\pm0.04$&$0.028\pm0.002$&$0.232\pm0.012$  &$9.80\pm1.61$&  $2.42\pm0.53$&$2.02\pm0.31$ &$5.30\pm0.01$&$1.25\pm0.01$&$4.05\pm0.01$&0.9/447 \\
57812.92441 &57813.49733 &$0.24\pm0.03$&$0.035\pm0.003$&$0.278\pm0.045$  &$10.81\pm2.02$& $2.23\pm0.32$&$1.35\pm0.22$ &$5.16\pm0.02$&$1.71\pm0.01$&$3.44\pm0.03$&1.1/446 \\
57839.91052 &57840.63969 &$0.06\pm0.01$&$0.051\pm0.013$&$0.346\pm0.031$  &$9.90\pm1.68 $&$1.72\pm0.13$&$0.42\pm0.03$ &$1.41\pm0.01$&$1.41\pm0.01$&$-$&1.0/311 \\
\hline
\end{tabular} 
\end{table*}


\subsection{X-ray variability}
To quantify the X-ray flux variation, we used the excess variance or the 
fractional variability 
amplitude $F_{\rm{var}}$ \citep{Edelsonetal1996,Nandraetal1997,2003MNRAS.345.1271V}. 
$F_{\rm{var}}$ is the variance after subtracting the contribution expected 
from measurement errors and is defined as 
\begin{equation}
\centering{\label{1}}
F_{\rm{var}}=\sqrt{{S^{2}-{\bar{\sigma^{2}}_{\rm{err}}}\over\bar{x}^2}}
\end{equation}
where ${\bar{x}^2}=\sum_{i=1}^{N}{x_i/N}$ is the arithmetic mean of $x_i$. ${S^{2}}$ and $\overline{\sigma_{\rm err}^{2}}$ are the sample variance of the light curve and mean square error associated with the measured fluxes $x_i$ respectively. 
\begin{equation}
\centering
S^2=\frac{1}{N-1}\sum_{i=1}^{N}(x_i-\bar{x})^2,
\end{equation}
\begin{equation}
\overline{\sigma_{\rm err}^{2}}=\frac{1}{N}\sum_{i=1}^{N}\sigma^{2}_{\rm{err,i}}
\end{equation}
Using Eq. ({\ref{1}}), we calculated the fractional variability amplitude for each segment of observations and the measurement uncertainties of $F_{\rm{var}}$ were estimated following \cite{2003MNRAS.345.1271V}
\begin{equation}
\centering{\label{2}}
\text{err}(F_{\rm{var}})=\sqrt{\Bigg(\sqrt{1\over{2N}}{\overline{\sigma_{\rm err}^{2}}\over{{\bar{x}}^2}{F_{\rm{var}}}}\Bigg)^2+\Bigg(\sqrt{\overline{\sigma_{\rm err}^{2}}\over{N}}{1\over\bar{x}}\Bigg)^2}
\end{equation}
The results of the variability analysis are listed in Table {\ref{table:Fvar}}. We found the amplitude of the flux variations to increase from epoch A to epoch D and 
the highest flux variations were seen in epoch D.
\begin{table}
\centering
\scriptsize
\caption{\label{table:Fvar} Flux variability characteristics for each
epoch of Mrk 421.}
\begin{tabular}{ccccc}   
\hline
OBS ID  & Epoch  & \multicolumn{3}{c} {$F_{\rm{var}} \pm \text{err}(F_{\rm{var}})$}  \\
        &  & 3$-$10 keV     & 10$-$20 keV     &  3$-$20 keV                     \\
\hline
60202048002  & A & 0.079$\pm$0.001  & 0.085$\pm$0.002 & 0.080$\pm$0.001   \\
60202048004  & B & 0.139$\pm$0.001  & 0.203$\pm$0.002 & 0.156$\pm$0.001 \\
60202048006  & C & 0.163$\pm$0.001  & 0.187$\pm$0.001 & 0.168$\pm$0.001 \\
60202048008  & D & 0.272$\pm$0.001  & 0.353$\pm$0.002 & 0.276$\pm$0.021  \\
\hline
\end{tabular} 
\end{table}

\begin{figure} 
    \centering{
    \includegraphics[scale=0.5,angle=0]{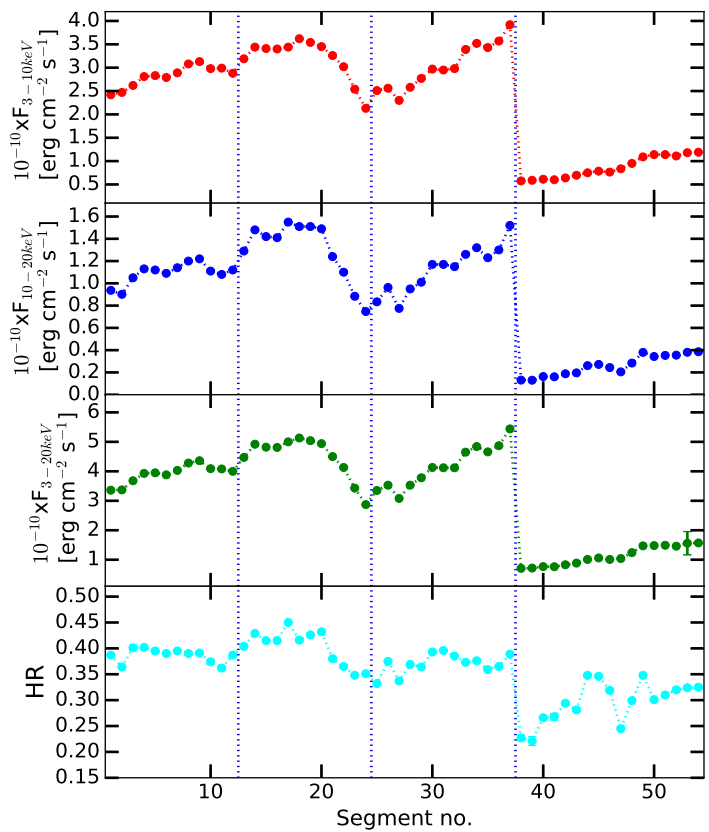}}
    \caption{Variation of the soft (3$-$10 keV), hard (10$-$20 keV) and total (3$-$20 keV) fluxes and hardness ratio (HR).
}
    \label{fig:HRtotflux}
\end{figure}

In the observed period presented here, the hardness ratio (HR) was calculated for each segment 
from the {\sc pl} model fitted spectra for the energy range 3$-$20~keV. In our 
case $HR=\frac{F_{10-20keV}}{F_{3-10 keV}}$, where, $F_{10-20keV}$ and 
$F_{3-10keV}$ are the flux in hard and soft energy bands. In 
\autoref{fig:HRtotflux}, we show the HR variation with total observed flux. 
In $\sim$83 days, HR changed by a factor of more than 2, and during the 
fourth epoch (2017 March 27), the source had the lowest hard flux.  
We performed the 
weighted least square fitting to the observed data points in the HR v/s flux 
diagram. We found a strong positive correlation between the HR and total flux. 
We obtained a correlation coefficient of 0.874 which also indicates that our 
spectrum become harder as the flux increases, confirming the results which are 
interpreted from photon index vs. flux diagram (\autoref{fig:PLindexFlux}).
It can also be seen from \autoref{table:segsPL} that there is a noticeable transition in flux in all energy bands. 
Between epoch C and epoch D the brightness in the soft band
reduced by a factor of 7 in magnitude, whereas in the hard band 
the brightness dropped  by a factor of 12. This large jump in flux between 
epoch C and D also implies that the corona which reprocesses soft photons 
from the Keplerian disc and scatters them as hard radiation has significantly 
cooled down, therefore can not produce enough hard radiation, and the source 
moved to lower flux state. This abrupt change in the flux is a signature of (a) sudden
change in flow configuration, which means viscosity has changed suddenly 
\citep{Mondaletal2017} or (b) appearance of some absorbers. 
The significant change in CENBOL geometry consistent with the flux 
variation can also be seen from the {\sc tcaf} model fitted 
parameters, where the size of the CENBOL has changed by a factor of $\sim 2$. 
Also, this is consistent with the estimated decrease in the temperature of the 
corona from epochs A to D.
From Table {\ref{table:Fvar}}, it is evident that hard X-rays are more 
variable than the soft X-rays. This might be indicative of more complex 
dependence of variability on energy and other physical mechanisms. The 
difference between soft and hard X-ray variability could be, for example, 
related to the size of the emitting region, or the corona and thus the 
cooling effect. There are arguments in the literature on the similar 
observational flux variability characteristics noticed here. If the 
high-energy electron cloud is located at the inner edge of the accretion disc, 
which is the Comptonizing corona in our case or above its inner part 
\citep{Zdziarskietal1999}, and the low-energy emitting region is associated to 
some outflow from the  disc or a second cloud  with  lower temperature 
\citep{Petruccietal2013}, then the hard flux could be more variable than the 
soft flux. This could explain both the harder when brighter trend in the
spectra and the hotter when brighter trend in the corona of Mrk\,421.

\subsection{Contribution of the jet to spectral flux} \label{sec:jetflux}
\begin{figure*}
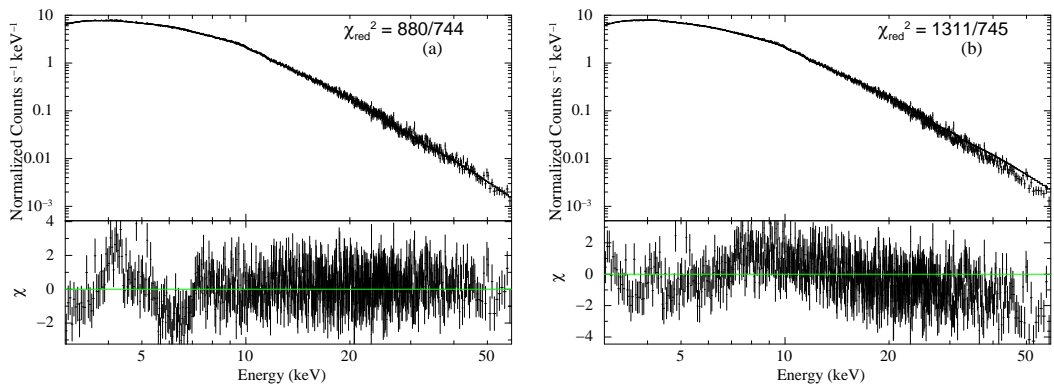
 
    \centering{
    \includegraphics[height=7truecm,angle=270]{Figures/spec-62332-tcaf-zxipcf-freeN-R1.eps}
    \includegraphics[height=7truecm,angle=270]{Figures/spec-62332-tcaf-zxipcf-FixedN-R1.eps}}
  \caption{The 3$-$60 keV spectra fitted with {\sc tbabs*zxipcf*tcaf} model along with the residuals for epoch E. The panels (a) shows the fit when {\sc tcaf} model norm is free and (b) corresponds to data fitted when norm is frozen to an average low flux state norm around 1.3e4.}
    \label{fig:specNorm}
\end{figure*}

In this section, we compare the spectral fitting with the accretion disc model to estimate the contribution from both disc and jet.
To investigate the relative importance of jet emission over the accretion
disc emission in Mrk\,421 as well as to compare the parameters obtained
from spectral fits to the data acquired during 2017,
we identified one {\it NuSTAR} data (OBSID = 60002023031, hereinafter Epoch E)
observed on 14 April 2013. The source was in a very active state in $\gamma$-rays,
optical and X-rays during epoch E, and the X-ray emission during this epoch is expected to 
be dominated by the emission from the relativistic jet. The R-band, the {\it Swift/XRT} soft and hard bands, and UV fluxes are 
$11.90\pm0.03$ (mag), $77.41\pm0.26$, $50.62\pm0.18$, and $30.97\pm0.74$ in ph\,s$^{-1}$ unit respectively. The hardness ratio (hard/soft flux) was 0.65 during this epoch.  
We fitted the spectra obtained during Epoch E, with {\sc pl}, {\sc thcomp} and {\sc tcaf}.
The {\sc pl} was not found to be useful, and indeed, returned a poor fit with $\chi_r^2$ of 3.7 during epoch E, while {\sc tcaf} fits well ($\chi_r^2$ of 1.18) the data with $\dot m_d=0.43$, $\dot m_h=0.304$, $X_s =27.47$, and $R=2.82$. The normalization from {\sc tcaf} fit was found to increase by a factor of about 60 times, compared to
the normalization obtained during epoch D, when the source was in the lowest flux state in
X-rays among all the epochs analysed here. The total X-ray flux was also high with the value $27.57\times 10^{-10}$ ergs~cm$^{-2}$~s$^{-1}$. In \autoref{fig:specNorm} we show the {\sc tcaf}
 model fitted spectrum when the normalization is free (in panel a) and for an average normalization (in panel b) obtained from the fit of other epochs. The panel (b) does not fit the data very well when the normalization for 
 the fit is frozen to the average value obtained from the relatively low flux data (epochs A-D) fitting. Moreover, it shows that the spectrum is softening gradually, this can be due to the cooling effect in jet and can be well explained by the {\sc logpar} model, often used for this source. Therefore, our comparative fitting of the less and very active jet epochs shows that the disc could be present, however, might be weak and subdominant. We also point out that multi-band data from UV/optical, X-ray, and $\gamma$ gathered during the same epochs analysed in this work was equally fit with synchrotron and SCC models \citep{MAGICmrk4212017data}.
  It is to be noted that for moderately emitting X-ray jets, 
{\sc tcaf} is enough to fit, since {\sc tcaf} uses cylindrical geometry for
interception flux calculations. For computation of average optical depth, {\sc tcaf} code does not
use a slab geometry given in \citet{SunyaevTitarchuk1980} rather it uses a spherical
geometry. In both the cases, the centrifugal force supported torus and matter inside the
funnel of the torus which constitutes the pre-jet matter till the sonic surface) 
were taken care of while computing the intercepted photons. For consistency, we checked that the {\sc thcomp} also fits the data well with $kT_{\rm e}=24\pm6$ keV and $\Gamma=2.41\pm0.05$.

As the normalization is a scale factor which depends on the source distance and its inclination, 
therefore it is constant between epochs for a particular source. If its value changes drastically, this could
point to an additional emission process contributing to the observed X-ray, for example, emission from the jet.
Therefore, the normalization can also be a probe to estimate the flux contributions 
from both the disc and jet. This has been successfully applied to black hole binary systems \citep[][and others]{Janaetal2017}. As Mrk\,421 is a jetted candidate, we applied that finding for 
this candidate as a case study. Considering that in the lowest normalization epoch, the flux contribution is mainly from the disc, and 
 increasing norm implies an excess contribution from the jet in addition to the disc contribution. Therefore, the total flux $F_{\rm total}$ can be written as the sum of disc 
and jet, as below:

\begin{equation}
    F_{\rm total} = F_{\rm disk} + F_{\rm jet}.
\end{equation}

\begin{figure*} 
    \centering{
    \includegraphics[height=8.0cm]{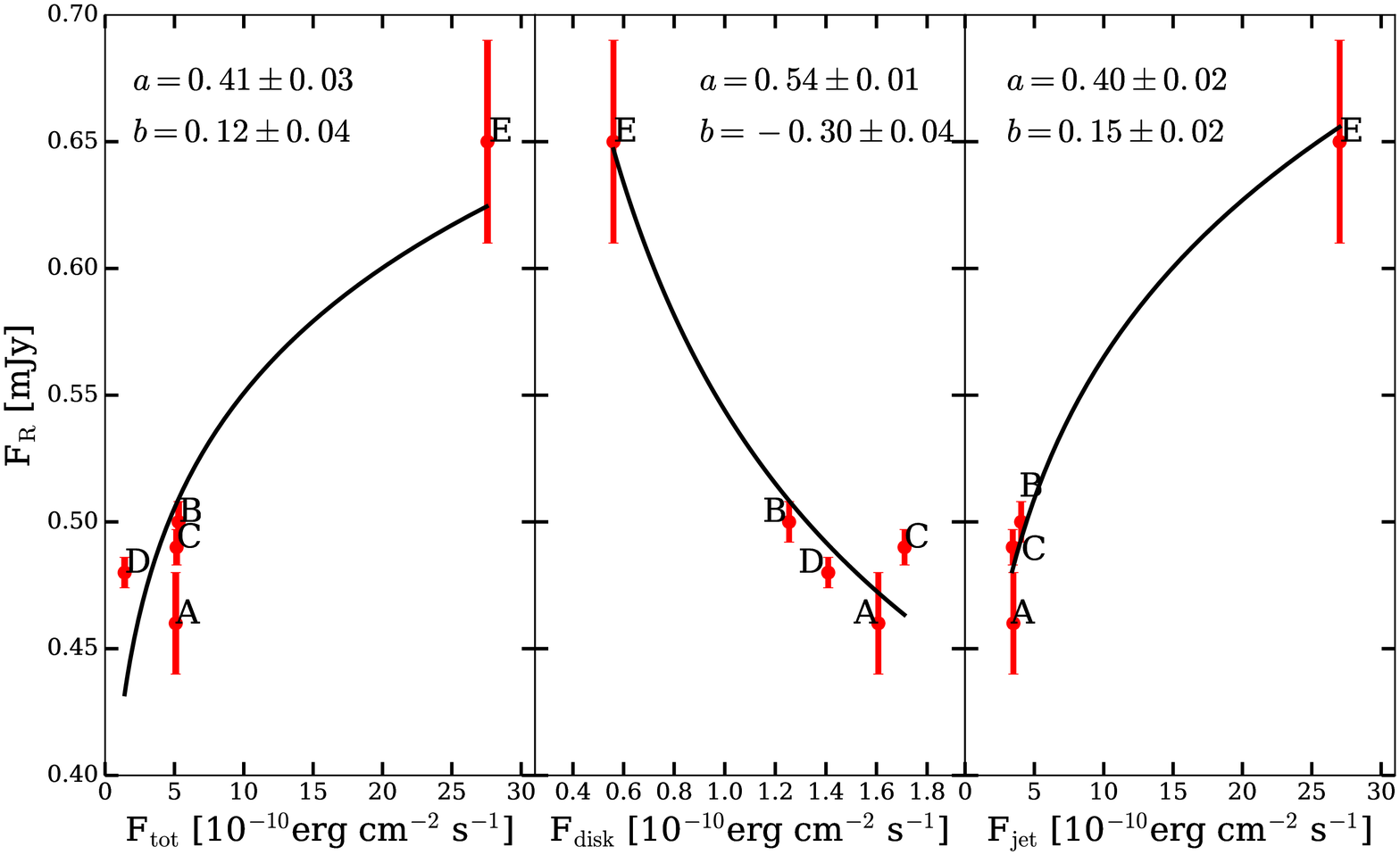}}
        \caption{Radio flux variation with different components of the flux obtained from {\sc tcaf} model fit. The red solid circles are the observed data points and the black solid line is the functional dependence between fluxes.} \label{fig:fluxes}
\end{figure*}

Here, $F_{\rm disk}$ can be obtained by using the lowest norm in epoch D to all other epochs and the jet contribution can be estimated by subtracting the lowest flux contribution from the disc from the total flux in \autoref{table:tcafPL} (in Column 8 when the norm is free). Therefore, for this epoch, F$_{\rm jet}$ is zero. This is also reflected in the third panel of \autoref{fig:fluxes}.
 
For all the epochs, we show in \autoref{fig:fluxes} the relation ($a x^b$) between the radio flux
(a proxy for jet emission) against the X-ray emission coming from
different components of the accretion-jet system. The three panels in \autoref{fig:fluxes} 
show the variation of the radio flux with the total flux, disc flux from
{\sc tcaf} fits and the jet flux obtained from Equation 5. In this figure, the red
circles are the observed/estimated flux values, while the solid black curve is
the fitted function. We find that the radio flux is anti-correlated with
the X-ray flux from the accretion disc, while it is correlated with the
jet X-ray emission. 
The normalization in {\sc tcaf} fits increases from Epochs D $\rightarrow$ C $\rightarrow$ A $\rightarrow$ B $\rightarrow$ E.
This indicates that the relative contribution of the jet emission over
the disc emission increases in the sequence D$\rightarrow$C$\rightarrow$A$\rightarrow$B$\rightarrow$E, with the X-ray jet emission
being maximum during epoch E and minimum during epoch D.

\subsection{Joint fit to the {\it Swift/XRT} and {\it NuSTAR} data} \label{sec:JointFit}
For estimating various accretion parameters as discussed in the previous section, we used only data from {\it NuSTAR} covering the energy range from 3-60 keV. It is likely, that inclusion of X-ray data with energies lower than 3 keV could have an impact on the derived accretion parameters. To test this, we did a joint spectral analysis of {\it Swift/XRT} and {\it NuSTAR} data with {\sc thcomp} and {\sc tcaf} models.

\begin{figure*}
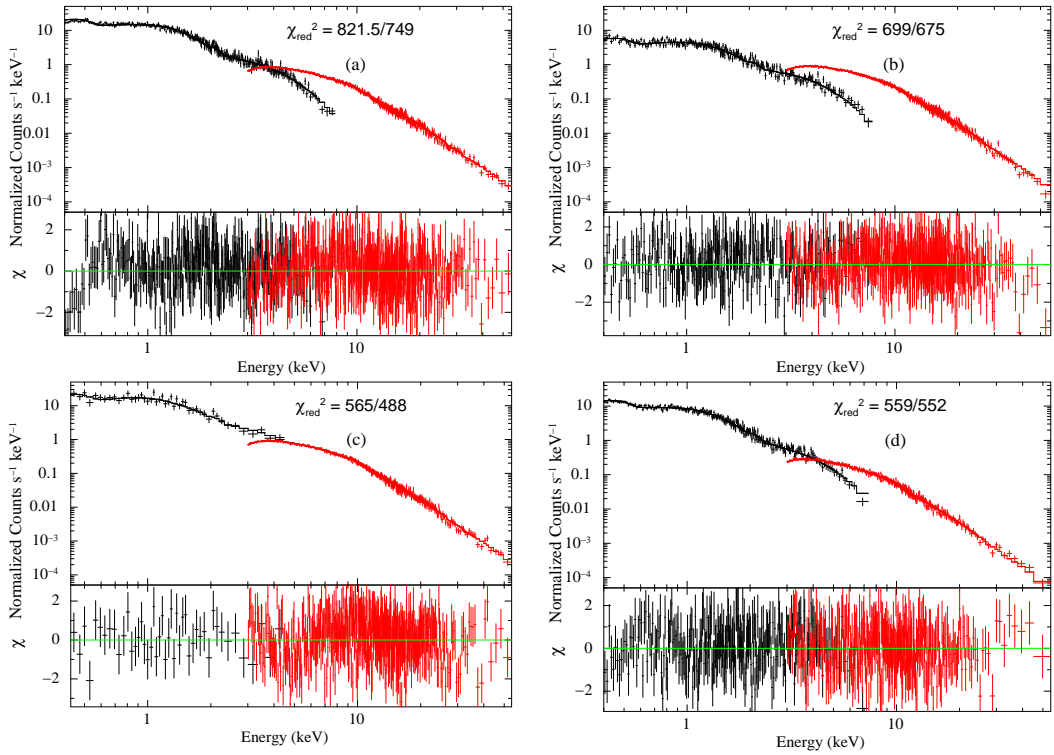
 
    \centering{
    \includegraphics[height=7truecm,angle=270]{Figures/x02-tcaf-gauss.eps}
    \includegraphics[height=7truecm,angle=270]{Figures/X04-tcaf-gauss.eps}}
  \centering{
  \includegraphics[height=7truecm,angle=270]{Figures/X06-tcaf-gauss.eps}
  \includegraphics[height=7truecm,angle=270]{Figures/X08-tcaf-gauss.eps}}
\caption{The 0.4$-$60 keV combined {\it Swift/XRT} and {\it NuSTAR} spectra fitted with {\sc tbabs*zxipcf*(tcaf+gauss)} model along with the residuals. The panels (a)-(d) correspond to the data acquired at epochs A, B, C and D respectively.}
    \label{fig:jointspectcaf}
\end{figure*}

For {\it Swift/XRT} data, the spectrum files were generated using the online product generator \citep{Evansetal2009}\footnote{https://www.swift.ac.uk/user\_objects/}. The data were binned to 20 counts rate per energy bin. For the joint spectral analysis, we used data in the energy range from 0.4 to 60 keV. For epochs A to D, {\it Swift/XRT} data covering the energy range of 0.4 to 8 keV was used except for
epoch C, where the data from energy range 0.4 to 5 keV was used, owing to poor S/N beyond 5 keV. The results of the {\sc thcomp} model fit to
the joint Swift/XRT and NuSTAR data are given in \autoref{table:jointThcomp}. A comparison of \autoref{table:jointThcomp} with that of \autoref{table-5} indicates the parameters obtained from combined {\it Swift/XRT} + {\it NuSTAR} data and {\it NuSTAR} data alone agree within errors. Similarly, we show in \autoref{fig:jointspectcaf}, the {\sc tcaf} model fits to the combined spectra, and the derived parameters are given in \autoref{table:jointTcaf}. In {\sc tcaf}
model fits too, the parameters obtained from the joint {\it Swift/XRT} and {\it NuSTAR} data agree within errors to the parameters obtained using only {\it NuSTAR} data. In model fits to the
combined data set, an extra Gaussian component was required to fit the excess flux in the 0.5 - 0.7 keV range. We note that the disc parameters (that are needed for the estimation of
viscosity parameter) obtained from joint fitting are in agreement to that obtained using {\it NuSTAR} data alone within errors.


\begin{table*}
\scriptsize
\centering
\caption{\label{table:jointThcomp} Results of  
{\sc tbabs*zxipcf*thcomp*(diskbb+gauss)} model fits to the joint {\it Swift/XRT} and {\it NuSTAR} spectra. The {\sc gauss} component is required at $\sim 0.5-0.7$ keV energy range to fit the excess around that energy. The other parameters are the same as in \autoref{table-5}.}
\begin{tabular}{cccccccccc}
\hline
Epoch &MJDstart &MJDend &$C_{\rm f}$  & $kTe$ &  $\Gamma$ & $T_{\rm in}$ &N$_{\rm dbb}$ &$\chi_r^2/dof$ \\
      & &&& [keV] &       &[keV] & $\times 10^{6}$ & \\
\hline
A &57756.99385 &57757.52163 & $0.57\pm0.11$ &$138\pm12$ &$2.40\pm0.01$& $0.07\pm0.01$ &$4.07\pm0.39$ &1.0/750\\
B &57784.99038 &57785.55288 & $0.63\pm0.03$ &$35\pm4$   &$2.43\pm0.02$&$0.07\pm0.02$ &$4.62\pm0.25$&1.0/676\\
C &57812.92441 &57813.49733 & $0.65\pm0.08$ &$38\pm5$   &$2.48\pm0.03$ &$0.07\pm0.02$&$3.91\pm0.40$&1.1/489\\
D &57839.91052 &57840.63969 & $0.12\pm0.01$ &$21\pm3$   &$2.54\pm0.04$ &$0.07\pm0.01$&$1.32\pm0.11$&1.1/553\\
\hline
\end{tabular} 
\end{table*}

\begin{table*}
\scriptsize
\centering
\caption{\label{table:jointTcaf} Results of {\sc tbabs*zxipcf*(tcaf+gauss)} model fits to the joint {\it Swift/XRT} and {\it NuSTAR} spectra. The {\sc gauss} component is required at $\sim 0.5-0.7$ keV energy range to fit the excess around that energy. The other parameters are the same as in \autoref{table:tcafPL}.}
\begin{tabular}{cccccccccccccccc}
\hline
MJDstart  &MJDend &$C_{\rm f}$&$\dot m_{\rm d}$ &$\dot m_{\rm h}$ & $X_{\rm s}$ &  R & $N_t\times 10^4$&$F_{\rm tot}$&$F_{\rm disk}$&$F_{\rm jet}$&$\chi_r^2/dof$ \\
\hline
57756.99385 &57757.52163 &$0.48\pm0.03$&$0.022\pm0.003$&$ 0.222\pm0.010$ &$20.09\pm2.48$& $2.41\pm0.36$&$0.99\pm0.13$&$14.23\pm0.04$&$5.42\pm0.02$&$8.81\pm0.05$&1.49/749 \\
57784.99038 &57785.55288 &$0.77\pm0.13$&$0.027\pm0.001$&$0.243\pm0.011$  &$9.91\pm1.23$ & $2.38\pm0.34$&$0.93\pm0.15$&$14.23\pm0.04$&$7.05\pm0.05$&$7.19\pm0.06$&1.0/675 \\
57812.92441 &57813.49733 &$0.78\pm0.16$&$0.034\pm0.004$&$0.292\pm0.011$  &$10.98\pm2.59$& $2.87\pm0.21$&$1.05\pm0.24$&$16.72\pm0.42$&$5.00\pm0.01$&$11.73\pm0.30$&1.15/488 \\
57839.91052 &57840.63969 &$0.46\pm0.02$&$0.051\pm0.008$&$0.348\pm0.021$  &$9.93\pm1.63 $& $1.73\pm0.19$ &$0.26\pm0.04$&$5.95\pm0.08$&$5.95\pm0.08$&--&1.0/552 \\
\hline
\end{tabular} 
\end{table*}

\subsection{Estimating accretion disc viscosity parameter}   
Flux variability of different time scales (hours to years) in AGN can be explained 
by several theoretical models (a) relativistic jet based models 
\citep{MarscherGear1985,GopalKrishnaWiita1992,CalafutWiita2015} where the variability
time scale is large and (b) 
the accretion disc based models where the variability can be in small scales. In case of the later models, it is believed 
that the observed X-ray variability in both black hole X-ray  binaries
and Seyfert galaxies mainly comes from two key components, accretion disc and 
its dynamic corona
where the X-rays are predominantly produced close to the BH by 
IC scattering of the optical/UV seed photons from 
the accretion disc. On the other hand, the geometrically thin, cold, and optically 
thick accretion disc reprocesses the variable X-ray emission coming from the 
innermost optically thin, hot corona in a complex way, therefore playing 
mostly a passive role in the variability at the shortest timescales. 
Furthermore, it has been interpreted by studying different timescales in 
accretion disc of blazars that the observed variations are solely due to jet 
or by the variations in the disc carried out, and amplified by 
jets \citep{Wiita2006}. Hence, the X-ray variability can be used to probe the 
properties of accretion on to the BH. The physical scenarios that describe the large scale 
variability in different accretion disc based models are the formation and fragmentation of spiral 
shocks in accretion disc \citep{Wiitaetal1992,ChakrabartiWiita1993}, the 
asymmetries and geometric effects in accretion disc  \citep{MangalamWiita1993},
or by the fluctuations that propagate from the outer radii of the accretion disc 
towards the center and couple with the emission from the inner region of the 
disc \citep{Lyubarskii1997,ArevaloUttley2006}.  It has been shown for black hole X-ray binaries 
that the viscosity and inverse Compton cooling can be responsible for small scale 
variability and different flux states \citep{Mondaletal2017} 
during the outburst phase of the BHs. The timescales of these variations can 
be scaled up for AGN through the BH mass \citep{McHardyetal2006}.

Studying the variability using any physical accretion disc model is important to understand the underlying dynamics behind these. X-ray flux variations in blazars in the low flux state can possibly be explained
by disc accretion. Very recently, \citet{Ritabanetal2018} discussed on the accretion disc origin of variability of Mrk\,421 using {\it AstroSat} data. Here, we use the results of the spectral analysis of the X-ray observations of Mrk\,421 by {\it NuSTAR} during 2017 using {\sc tcaf} model to understand the flux variations possibly being caused by accretion disc through the accretion rate variation, which also represents the viscosity ($\alpha$) fluctuation. 

It is believed that $\alpha$ parameterization provides the basis for developing the accretion disc theory and related observations. There are several methods to estimate $\alpha$ parameter. For instance, in the case of observations, if we assume that the optical variability of the AGN is caused by disc instabilities, then comparing the thermal timescales of $\alpha$-disc \citep[][hereafter SS73]{Shakura1973} models with the observed variability timescales, one can put constraints on the viscosity parameter for different candidates \citep{SiemiginowskaCzerny1989,Starlingetal2004}. In the numerical simulation, \citet{Balbus1991} showed that $\alpha$-parameter can be obtained from the outward transport of angular momentum in weakly magnetized disc by magnetohydrodynamic turbulence. \citet{Pessahetal2007} inferred observationally that in order to make MRI-driven turbulence, the angular momentum transport is required for large values of the effective 
$\alpha$ viscosity ($\alpha \ge 0.1$), the disc must be threaded by a significant vertical magnetic field and the turbulent magnetic energy must be in near equipartition with the thermal energy. On the other hand, in an advection-dominated disc,  \citet{NarayanMcClintock2008} argued that the required viscosity is 
0.1$-$0.3. All these  estimates seem to show a very broad range of the $\alpha$ parameter, 0.001$-$0.6 \citep[see also][]{Hawleyetal1995}. However, a narrow range of $\alpha$ parameter for blazars has been estimated by \citet{Xieetal2009}. In this work, we made an attempt to estimate the viscosity parameter for Mrk\,421.

To estimate the viscosity parameter we used the  disc  accretion 
rate returned by {\sc tcaf} model fits to the spectra. We assumed that the disc 
radiates locally as a black body with an effective temperature,
$$
T=\left(\frac{3GM_{\rm BH}\dot M}{8\pi\sigma}\right)^{1/4} f^{1/4} r^{-3/4},
$$
where $f=[1-(r_{\rm in}/r)^{1/2}] \sim 1$, $r$ is the radial distance of the disc. We considered the outer boundary 
of the disc as 500~$r_s$. The kinematic viscosity $(\nu)$ can be written as
$$
\nu=\frac{\dot M}{3\pi \Sigma},
$$
where, $\Sigma$ is the surface density of the disc and can be obtained from 
the density ($\rho$ in gm~cm$^{-3}$) and scale height ($H$ in cm) of the disc. According to SS73 prescription, $\nu$ can be written as $\alpha c_s H$, where $c_s$ is the isothermal sound speed, corresponding to the effective disc temperature. After combining these equations and performing a few steps of algebra, we obtained the $\alpha$-parameter value. Our estimated values of $\alpha$ ranges 
between 0.18$-$0.25 in the Keplerian component of the two component disc. We also estimated different time scales of the flow, 
for example the dynamical ($t_{\rm dyn}=\sqrt{GM_{\rm BH}/r^3}$) and thermal 
time scales ($t_{\rm th}=\alpha^{-1}t_{\rm dyn}$) are $\sim$ 2 years and 
8$-$10~years respectively, whereas the viscous timescale 
($t_{\rm vis}=t_{\rm th}(H/r)^{2}$) is $\sim$ 30$-$36~Myr. 

\section{Conclusions}
In this paper, we studied the flux and spectral variability of Mrk~421 using 
{\it NuSTAR} data obtained in 2017 and in 2013. 
During the 2017 epoch, we found the {\sc pl} photon index to change from 
2.2 to 3.0. We split each epoch of observation into several segments to 
study the flux and spectral variability on short time scales. We noticed
that while the hardness ratio (\autoref{table:segsPL}) is almost doubled, 
the average count rate of the source varied 
by a factor of 4 in $\sim 83$ days. We found that each epoch of
observation in the energy band 3$-$60~keV could be well fitted with phenomenological
{\sc thcomp} and physical {\sc tcaf} models. Each segment of data 
could be well fitted with the simple {\sc pl} model. From the obtained 
$F_{\rm var}$ (in \autoref{table:Fvar}), we conclude that the source was 
significantly variable in all the epochs. We also found a strong correlation 
between the {\sc pl} index and the brightness of the source in the total energy
band with a ``harder when brighter" behaviour.

Mrk\,421 is a HSP blazar which generally lacks emission lines in the optical spectra. 
Under such circumstances, the origin of the
observed X-ray flux variations lie in their jets 
\citep{Abdoetal2011}. 
Alternatively, during the less active state, 
one may expect the contribution of accretion disc to the observed emission, however, unlikely to dominate over jet contribution.
Recently, \cite{Ritabanetal2018} noticed a break in the power spectral
density from X-ray observations of Mrk\,421 at around the same time
 as that of the 
{\it NuSTAR} data which is analysed here. The observations from {\it AstroSat} used 
by \cite{Ritabanetal2018} were also taken during the moderate brightness state of 
Mrk\,421. Such breaks, that are normally
seen in the X-ray power spectral density of Seyfert galaxies, are believed to be due to 
the X-ray being produced in the accretion disc. 
We therefore, hypothesise that in the low or moderate X-ray activity of Mrk\,421, disc emission could manifest itself in the observed X-rays and thus fit each epoch of data using the disc-based {\sc tcaf} model to investigate if the observed flux variations can be explained with changes in the accretion rate or changes in the geometry of the accretion disc.
The derived parameters from the model fits show that the disc mass accretion rate has
varied by a factor of $\sim$3 from $0.02$ to $0.051$ $\dot M_{\rm Edd}$ and the size 
of the CENBOL as envisaged in {\sc tcaf} model 
also changed significantly. This size varied 
from 20 to 10~$r_s$ between epochs A and D. In 
epoch D, the inner edge of the disc moved significantly
inward with the lowest shock compression ratio and the shrinking of 
the CENBOL was maximum, and the accretion rate was the 
highest. This implies a transition of the source to a 
lower flux state with a significant flux change. Also, the
temperature of the CENBOL was found to change from epoch A to D, using both
physical {\sc tcaf} and phenomenological {\sc thcomp} model fits. 
The CENBOL was found to be hotter with increasing brightness of the source. 
Therefore, the shrinking of the CENBOL, 
increase of the mass accretion rates, and the decrease in the flux values 
follow the correlation with $\Gamma_{\rm PL}$ variation. Even though
both the model temperatures are showing the same profile, the static corona (used in {\sc thcomp})
cannot behave in the same way as the dynamic corona/CENBOL (used in {\sc tcaf}) 
does because the static corona would cool quickly if there is no underlying heating mechanism. Therefore, the temperature estimated from {\sc tcaf} model is more realistic. 

We note that the accretion disc-based model like {\sc tcaf} can fit the spectra without 
adding any other component. This indicates that even though 
Mrk\,421 is a HSP blazar, in the low/moderate activity state during 
the period of our observation, the disc could contribute significantly to the total observed X-ray emission, however,
not dominating over the jet contribution.
The difference in the flux variations between the hard and soft bands
and subsequently the hardness ratio could also be due to changes in the
size of the emission region or the CENBOL. 
Thus, the hard emission might have its origin from the changing 
corona region at the inner edge of the disc. On the other 
hand, the presence of jet and its contribution to the total spectra 
would imply that the gravitational potential energy of 
the infalling matter not only gets transformed into radiation, but can also 
amplify magnetic field, that allows the field to retrieve large store of 
rotational energy and transform a part of it to jet power.

Thus, based on the modelling of the five epochs of {\it NuSTAR} data of Mrk\,421 for the energy band 3-60~keV from the relatively less and very high jet activity, 
we conclude that (i) simple {\sc pl} alone is not a good 
representation of the observed X-ray rather {\sc cutoffpl} fits the data well, (ii) both {\sc tcaf} and 
{\sc thcomp} fit the observed X-ray emission well 
to give the accretion and spectral properties though {\sc tcaf} gives the 
physical parameters of the flow, 
(iii) non-magnetic accretion disc models are found to be adequate to fit the low/moderate X-ray activity state data of Mrk\,421. However, the X-ray/$\gamma$-ray correlation is one of the important characteristics in the broadband emission of Mrk\,421, that occurs during high and low activity, as reported multiple times in the literature over the last years, and particularly for the data obtained in the year 2017, as reported in \citet{MAGICmrk4212017data}. The existence of this correlation implies a substantial emission from the jet, probably related to SSC models, that occurs also during the very low activity. And hence that, even during the low blazar activity, the present disc-base models can not dominate the X-ray emission of Mrk\,421.

\begin{acknowledgements}
We thank the referee for the critical comments on our manuscript. SM acknowledges Andrzej Zdziarski for helpful discussions and helping in model fitting.
SM thanks Keith A. Arnaud for helping in model implementation in XSPEC package. 
SM acknowledges funding from Ramanujan Fellowship grant (\# RJF/2020/000113) by SERB-DST, Govt. of India.
This research has made use of the {\it NuSTAR} Data Analysis Software ({\sc nustardas}) jointly developed by the ASI Science Data Center (ASDC), Italy and the California Institute of Technology (Caltech), USA. This research has also made use of data obtained through the High Energy Astrophysics Science Archive Research Center Online Service, provided by NASA/Goddard Space Flight Center.

\end{acknowledgements}

\bibliography{variability}{}
\bibliographystyle{aa}

\end{document}